\RequirePackage{etex} 
\documentclass[aps,prb,english,longbibliography,twocolumn,superscriptaddress]{revtex4-2}
\usepackage[active]{srcltx}
\usepackage[utf8]{inputenc}
\usepackage{latexsym}
\usepackage[colorlinks=true,linkcolor=blue,citecolor=red]{hyperref} %
\usepackage{orcidlink}
\usepackage{amsmath}
\usepackage{xcolor}
\usepackage{amssymb}
\usepackage{amsfonts}
\usepackage{graphicx}
\usepackage{slashed}
\usepackage{wasysym}
\usepackage{pifont}
\usepackage{manfnt}
\usepackage{tikz}
\usepackage{tikz-3dplot}
\usepackage{tikz-network}
\DeclareSymbolFont{usualmathcal}{OMS}{cmsy}{m}{n}
\DeclareSymbolFontAlphabet{\mathcal}{usualmathcal}
\usepackage{braket}
\usepackage{subfig}
\usepackage{ragged2e}

\newcommand{\be}{\begin{equation}}
	\newcommand{\ee}{\end{equation}}
\newcommand{\bea}{\begin{eqnarray}}
	\newcommand{\eea}{\end{eqnarray}}

\newcommand{\mc}{\mathcal}

\begin{document}
\title{Quantum many-body scarring from Kramers-Wannier duality}
\author{Weslei B. Fontana \orcidlink{0000-0001-8368-5997}}
\affiliation{Department of physics, National Tsing Hua University, Hsinchu 30013, Taiwan}
\author{Fabrizio G. Oliviero \orcidlink{0000-0002-5721-7687}}
\affiliation{Department of physics, National Tsing Hua University, Hsinchu 30013, Taiwan}
\affiliation{Physics Division, National Center for Theoretical Sciences, Taipei 10617, Taiwan}
\author{Yi-Ping Huang\,\orcidlink{0000-0001-6453-1191}}
\email[Correspondence email address: ]{yphuang@phys.nthu.edu.tw} 
\affiliation{Department of physics, National Tsing Hua University, Hsinchu 30013, Taiwan}
\affiliation{Physics Division, National Center for Theoretical Sciences, Taipei 10617, Taiwan}
\affiliation{Institute of Physics, Academia Sinica, Taipei 115, Taiwan}
\begin{abstract}
    Kramers-Wannier duality, a hallmark of the Ising model, has recently gained renewed interest through its reinterpretation as a non-invertible symmetry with a state-level action. Using sequential quantum circuits (SQC), we argue that this duality governs the stability of quantum many-body scar (QMBS) states in a nonintegrable model, depending on whether the dual preserves the embedding conditions for scarring. This is supported by good agreement between first-order perturbation theory and numerics, which capture scar dynamics despite chaotic spectra. Our results establish non-invertible dualities as both a generative mechanism and a diagnostic tool for quantum many-body scarring, offering a generalized symmetry-based route to weak ergodicity breaking.
\end{abstract}
\maketitle

\section{Introduction}Understanding or preventing thermalization in isolated quantum many-body systems is a central challenge in modern physics. While most such systems are expected to thermalize in accordance with the Eigenstate Thermalization Hypothesis (ETH) \cite{Berry1977,deutsch1991,Srednicki1994}, notable exceptions have emerged \cite{Gogolin2016, Huse2015, Alet2018, Serbyn2019}. Among them, quantum many-body scars (QMBS) form a distinct class: nonthermal eigenstates embedded within otherwise thermal spectra, which were first observed in synthetic quantum systems \cite{Bernien2017}, and have since been investigated through a growing body of numerical and analytical studies \cite{Turner2018,Iadecola2019,Lin2019,odea2020}. Their discovery has sparked intense interest due to both their fundamental significance \cite{Serbyn2021,papic2022weak,Moudgalya2022, Chandran2023} and potential for quantum applications \cite{dooley2021,dooley2023}.

Despite significant progress from both fundamental and applied perspectives, the generic mechanisms to stabilize QMBS remains an intriguing and open question. In contrast to ground states, which can be stabilized by energy gaps or symmetry protection—as in topological phases—scar states lie at finite energy densities and typically lack such protection, rendering them fragile to hybridization with thermal states \cite{Motrunich2020,Surace2021, Dalmonte2021}. This raises a central question: Can there exist structural principles—beyond symmetry or integrability—that characterize, constrain, or even protect non-thermal subspaces in chaotic quantum systems? While numerous mechanisms have been proposed to realize QMBS \cite{Shiraishi2017, Chamon2019, Mougdalya2020, odea2020, Moudgalya2024,TaoLin_2025, ben2025many, nicolau2025fragmentation, jonay2025localized, bhowmick2025granovskii, halder2025}, the role of \emph{duality} remains underexplored. Dualities are a cornerstone in understanding ground-state phases, but their application to highly excited states has been limited-- largely due to the challenge of tracking their action on individual eigenstates. This conceptual gap persists despite intense interest in the microscopic structure and robustness of QMBS.

In this work, we study the role of duality transformations in the stability of quantum many-body scars. We focus on a Hamiltonian constructed via the stochastic matrix form (SMF) decomposition \cite{Castelnovo2005, Siggia1977}, which hosts exact scar states at zero energy \cite{Chamon2019}. Using the Kramers-Wannier (KW) duality as a test case, we present two key findings:
(i) Certain scar states retain their nonthermal structure under duality, providing a constructive mechanism for generating new QMBS models.
(ii) Other scar states are mapped into thermal-like eigenstates by duality, enforcing ETH behavior and highlighting intrinsic fragility.
These results suggest that duality acts as a structural lens to identify and assess the stability of corresponding quantum many-body scars. To investigate duality at the level of many-body eigenstates, we employ sequential quantum circuits (SQCs) \cite{Wolf2005, Wolf2008, Solano2007}, which implement global dualities through local unitary gates. SQCs tracks duality actions at the level of many-body eigenstates \cite{Xie2023, Xie2024,Shao2024, Seifnashri2024} and naturally preserve area-law entanglement and admit matrix product operator (MPO) representations of finite bond dimension \cite{Tantivasadakarn2024, Verstraete2023}, making them ideally suited for analyzing scar states—whose entanglement structure is typically far from volume-law. This framework enables us to track duality mappings explicitly within the spectrum and access excited-state structure beyond what conventional operator-based dualities offer.

This work is organized as follows: In section \ref{sec: Model} we introduce our model that consists of a combination of two fixed point SMF Hamiltonians that define the limits in which we have exact isolated scars. In one of the limits, we show that it is possible to obtain a unique scar state which is built on top of a trivial product state. The form of the state is simple enough to compute its correlations analytically. The other part of the model, describes a degenerate scar subspace akin to the spontaneous symmetry breaking of the $\mathbb{Z}_2$ symmetry of the model. It can be shown that the state is a simple product state, for which all properties can be obtained systematically. In Section \ref{sec: Duality} we introduce the Kramers-Wannier duality operator, that is responsible for mapping between symmetric scar subspaces, and further is an indicative of a phase transition between the ground states of the model at the self-dual point, for which the operator is enhanced to a non-invertible symmetry. In section \ref{sec: scars signatures} we present results from exact diagonalization simulations to provide further evidence that the QMBS states have a indicative of robustness, in particular we discuss how the computation of matrix elements of the perturbing interactions signal non-thermal behavior in case the scars states are perfectly mapped by the duality operator. Moreover, in \ref{sec: fidelity loss} we complement the study of stability computing the fidelity of the scars states against perturbing terms and compare with the prediction of perturbation theory. Finally, in section \ref{conclusions} we end with some comments on the validity of our methods and some open questions for future investigation.
\section{Model\label{sec: Model}}
We consider $N$ qubits placed along the sites of an one-dimensional chain, their interactions are governed by the Hamiltonian $H=J_1H_{\text{D}}+J_2 H_{\text{ND}}$, where each term is defined as
\begin{align}
    H_\text{D}(\beta) = \sum_{j=1}^{N-2} \alpha_j\mc Q_j +\alpha_{N-1}\mc Q_{N-1}
    +\alpha_N\mc Q_N^{(t)}\,,\label{eq: hamiltonian degenerate}\\
    H_{\text{ND}}(\beta)=\sum_{j=1}^{N-2}\alpha_j\bar{\mc Q}_j+\alpha_{N-1}\bar{\mc Q}^{(t')}_{N-1}+\alpha_N\bar{\mc Q}^{(t')}_N\,,
    \label{eq: hamiltonian non degenerate}
\end{align}
here we assume staggered couplings $\alpha_j =\alpha+(-1)^j$ with $|\alpha|<1$ and $X_j,\,Z_j$ stand for the corresponding Pauli matrices at site $j$. We define $\mc Q,\,\bar{\mc Q}$ as
\begin{align}
    \mc Q_j&=\exp{\left[-\beta\left(X_j+X_{j+1}\right)\right]}-Z_jZ_{j+1}\,,\\
    \mc Q_N^{(t)} &= \exp{\left[-\beta\left(X_N+X_1\right)\right]}-tZ_NZ_1\,,\\
    \bar{\mc Q}_j&=\exp{\left[-\beta\left(Z_{j}Z_{j+1}+Z_{j+1}Z_{j+2}\right)\right]}-X_{j+1}\,,\\
    \bar{\mc Q}^{(t')}_{N-1}&=\exp{\left[-\beta\left(Z_{N-1}Z_N+t'Z_NZ_1\right)\right]}-X_{N}\,,\\
    \bar{\mc Q}^{(t')}_{N}&=\exp{\left[-\beta\left(t'Z_{N}Z_1+Z_1Z_2\right)\right]}-X_1\,,
\end{align}
where parameters $t,\,t'\in \mathbb{Z}_2$ label twist sectors. This model has a global spin-flip symmetry implemented by $\eta = \prod_{j=1}^N X_j$, and for finite and even $N$, it preserves lattice reflection by the center of the chain. The total Hilbert space is labeled by the pair $\left(\text{symmetry}, \text{twist}\right)$.

\begin{figure}
    \centering
    \subfloat[]{\includegraphics[width=\linewidth]{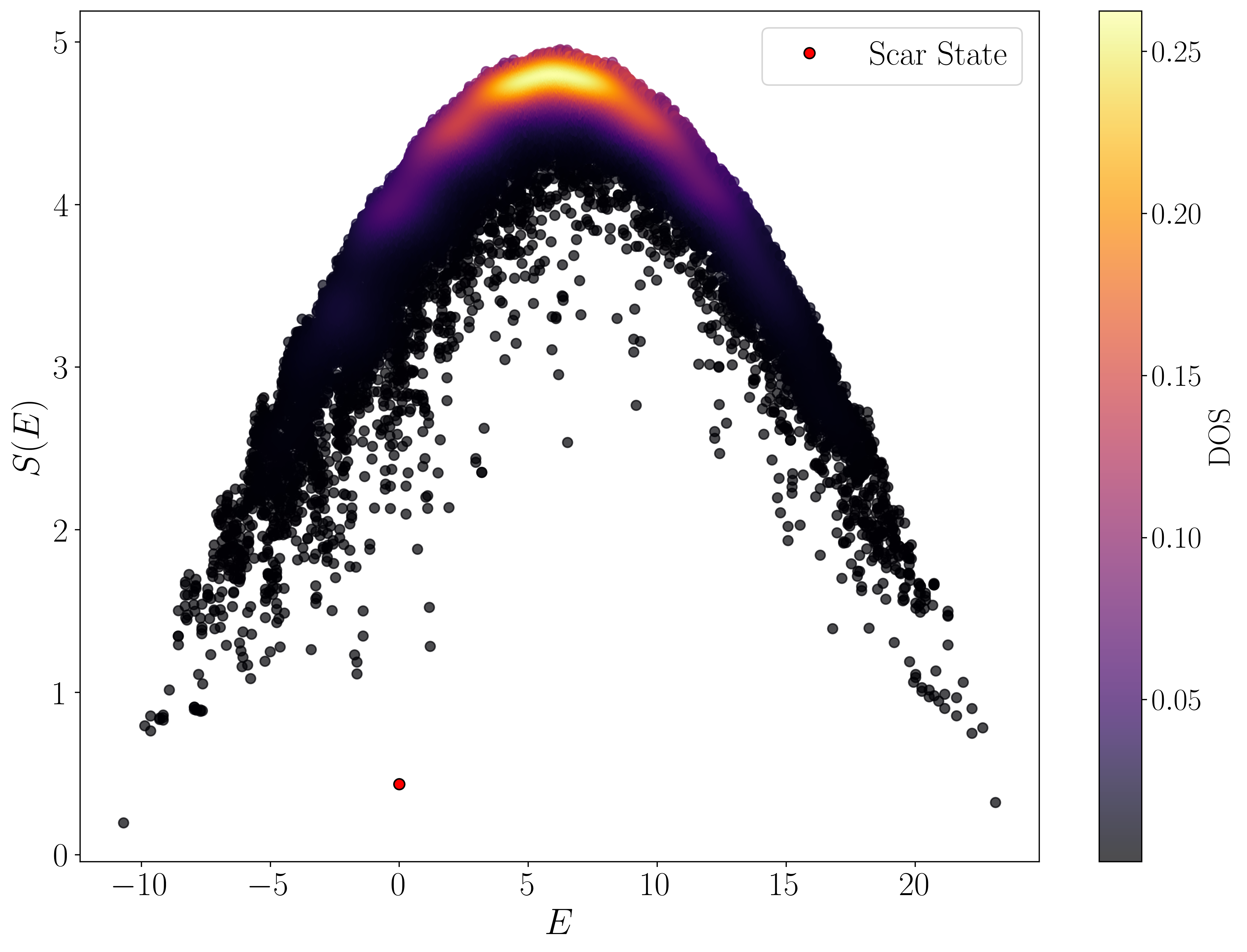}}\\
    \subfloat[]{\includegraphics[width=\linewidth, scale=0.3]{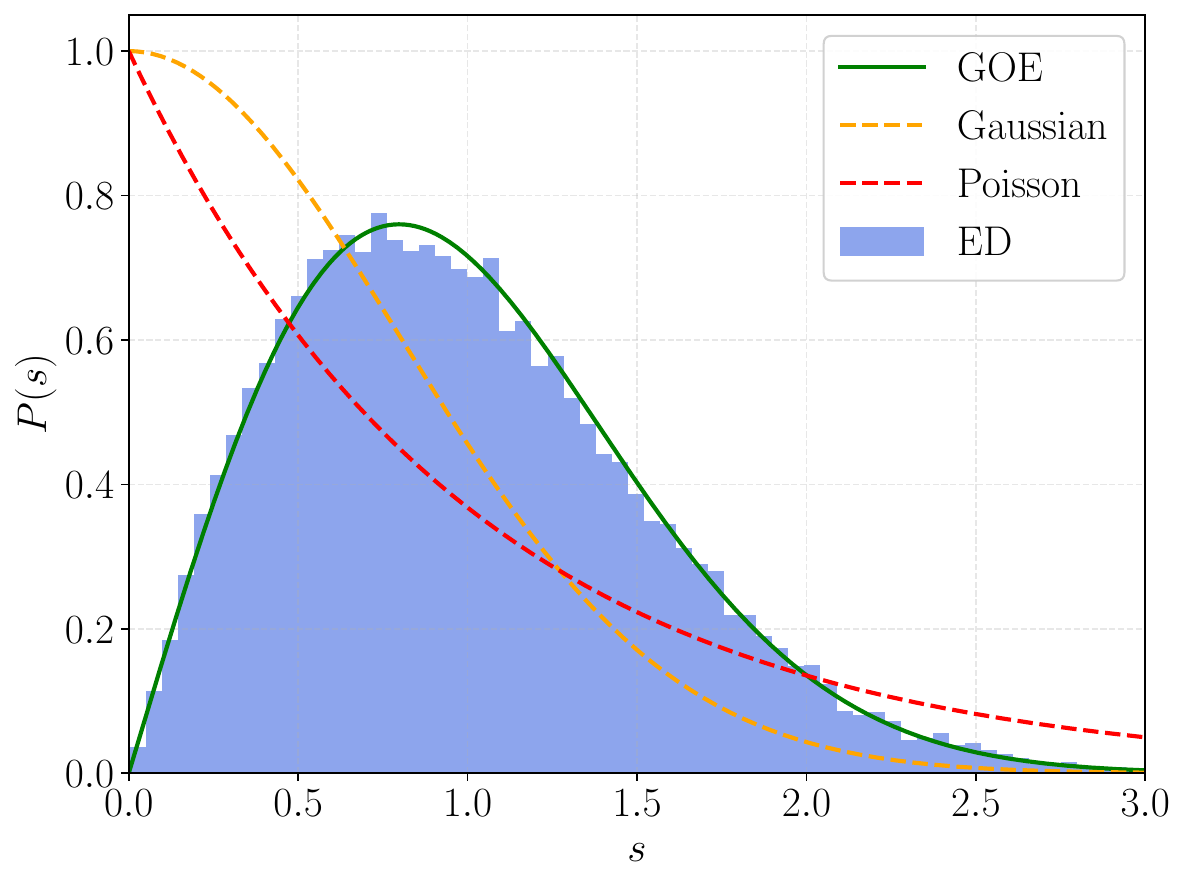}}
    \caption{\justifying (a) Entanglement entropy versus energy obtained from an exact diagonalization procedure for a chain of size $L=16$ for the $H_{\text{ND}}$ Hamiltonian. The scar state correspond to the isolated state at zero energy at low entropy in comparison to the rest of the spectrum. (b) Level statistics of paramagnetic Hamiltonian with $\alpha=0.3$ and $\beta=0.5$. Obtained from exact diagonalization for a chain of size $L=20$ sites. The plot is obtained by using the middle $\sim70\%$ of the spectrum in each momentum sector excluding $k=0,\,\pi$.}
    \label{fig:entanglement-para-scar+level-spacing}
\end{figure}

\begin{figure}
    \centering
    \subfloat[]{\includegraphics[width=\linewidth]{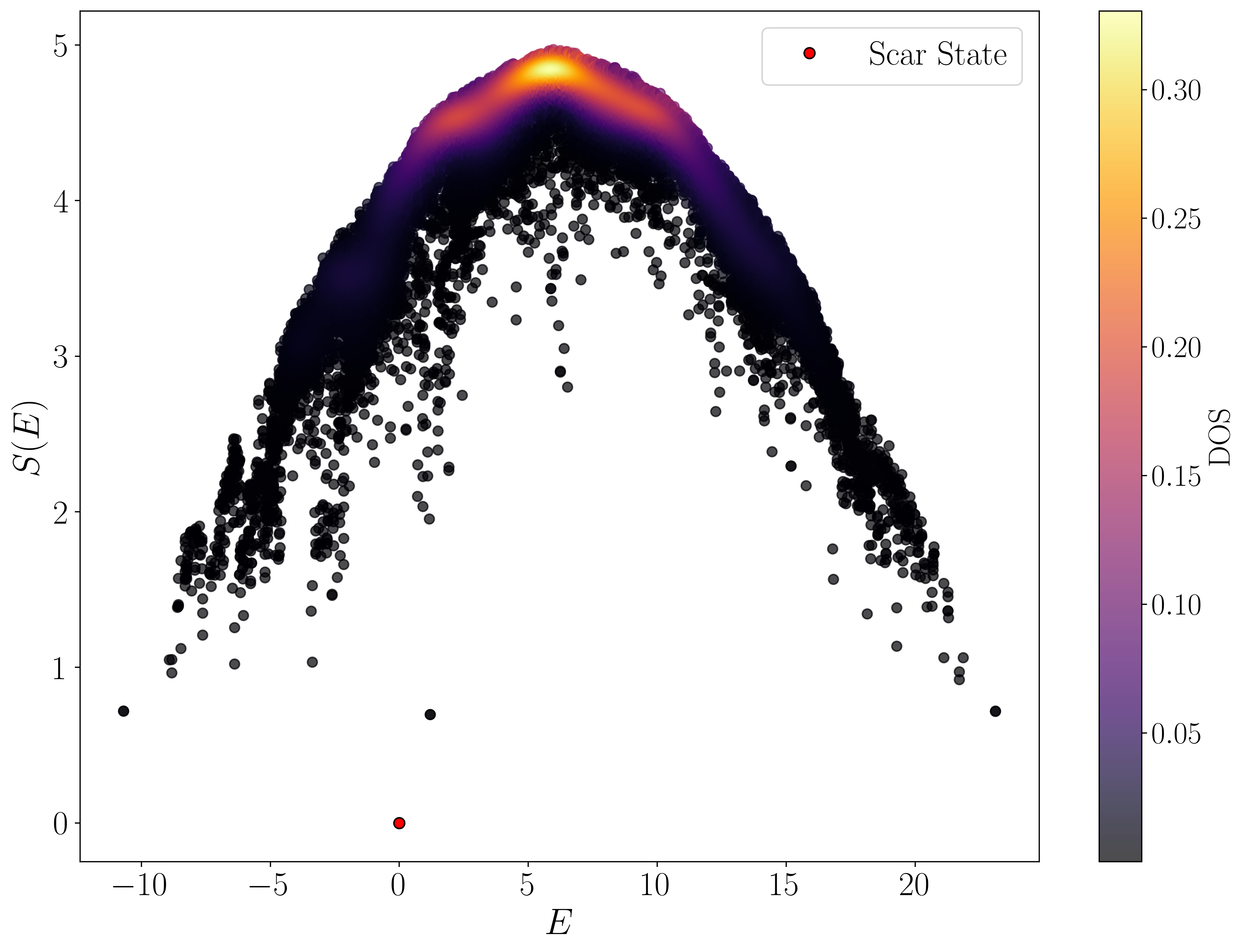}}\\
        \subfloat[]{\includegraphics[width=\linewidth]{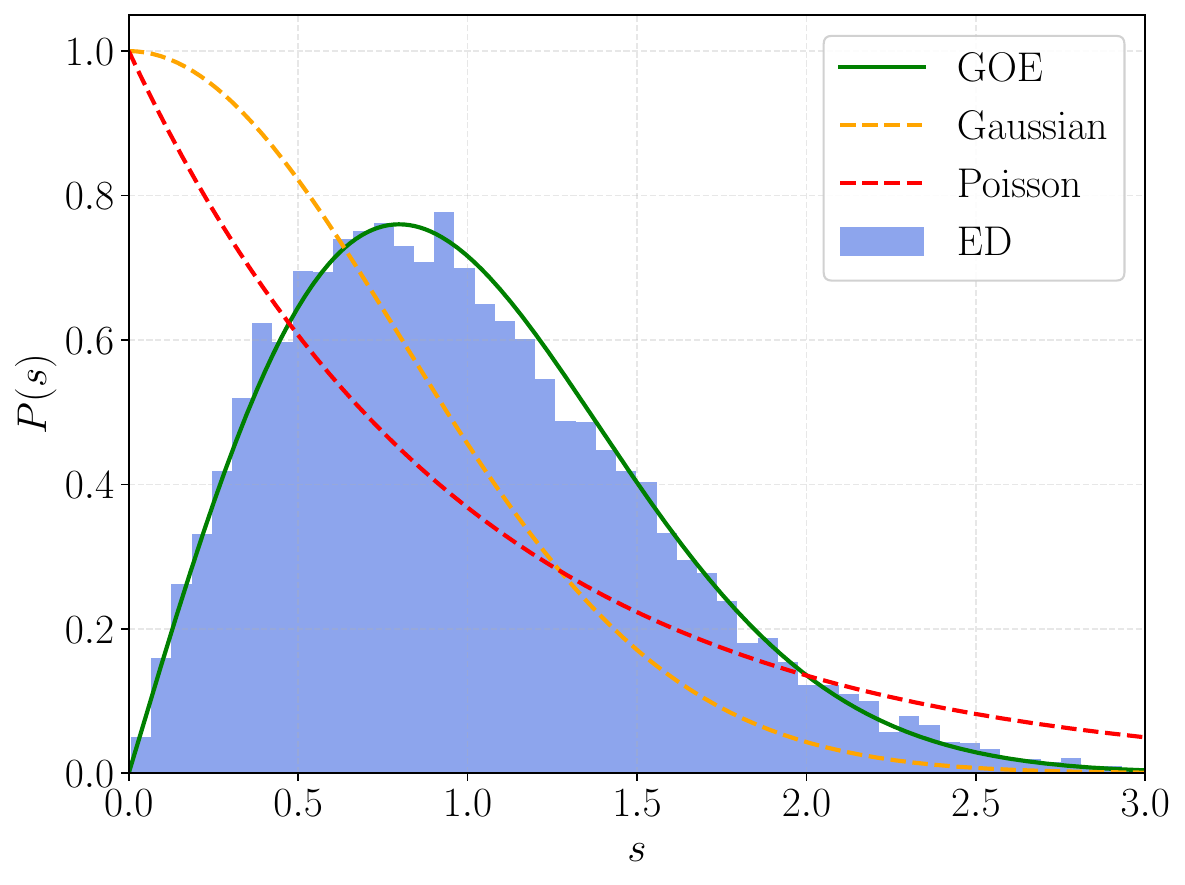}}
    \caption{\justifying{(a) Entanglement entropy versus energy obtained from an exact diagonalization procedure for a chain of size $L=16$ for the $H_\text{D}$ Hamiltonian. The two degenerate scar states sit at exactly zero energy and zero entropy, due to the product state form of the wavefunction. (b) Level statistics of nondegenerate Hamiltonian  with $\alpha=0.3$ and $\beta=0.5$. Obtained from exact diagonalization for a chain of size $L=20$ sites. The plot is obtained by using the middle $\sim70\%$ of the spectrum in each momentum sector excluding $k=0,\,\pi$.}}
    \label{fig:entanglement-ferro-scar+lvlspacingferro}
\end{figure}

Each Hamiltonian $H_{\text{D}}$ and $H_{\text{ND}}$, when considered separately, can admit QMBS states in their untwisted sectors, i.e., $t,\,t'=+1$. In this case, each Hamiltonian admits a SMF structure \cite{Castelnovo2005}. This class of systems have very interesting properties; they are composed of positive semi-definite terms of the form $\mc Q_s(\beta) = \exp(-\beta M_s)-A_s$, with $\left\{A_s,\,M_s\right\}=0$, $A_s^2=\openone$, which implies $Q_s^2=2\cosh\left({\beta M_s}\right)Q_s$. 
For non-vanishing $\beta$, one can show that each part of $H$ is non-integrable from its level spacing statistics, that obey a Wigner-Dyson distribution \cite{Huse2007, Santos2010, Bogomolny2013, Rigol2010, Collura2012}. We perform exact diagonalization for a chain of size $N = 20$ and obtain that the average of level spacings is $\langle r_ {\text{D}}\rangle=0.5299$, and $\langle r_ {\text{ND}}\rangle=0.5303$, which aligns well with the GOE value  $\langle r_ {\text{GOE}}\rangle=0.5359$, and is clearly distinct from the Poisson value $\langle r_ {\text{Poisson}}\rangle=0.3863$. The level statistics and entanglement profiles of the Hamiltonians $H_{ND}$ and $H_D$ within their scarred sectors can be seen at Fig.\ref{fig:entanglement-para-scar+level-spacing} and Fig.\ref{fig:entanglement-ferro-scar+lvlspacingferro}, respectively.

At the fixed point regimes, $J_1=0$ (or $J_2=0)$ the model admits exact QMBS states that are annihilated by the Hamiltonian as $H_\text{D}(\beta)\ket{\mc S_a(\beta),\,t=+1}=0$, with
\begin{equation}
    \ket{\mc S_a(\beta),\, t=+1} = e^{\frac{\beta}{2}\sum_i X_i}\otimes_i \ket{a}_i\,, ~~~a=\left\{0,1\right\}\,,
    \label{eq: ferro scars}
\end{equation}
and $H_{\text{ND}}(\beta)\ket{\bar{\mc S}(\beta),\,t'=+1}=0$, with
\begin{equation}
    \ket{\bar{\mc S}(\beta),\, t'=+1} = e^{\frac{\beta}{2}\sum_i Z_iZ_{i+1}}\otimes_{j}\ket{+}_j\,.
    \label{eq: scar state}
\end{equation}

In our convention $\ket{+}=\frac{1}{\sqrt{2}}\left(\ket{0}+\ket{1}\right)$, with $\ket{0}$ and $\ket{1}$ being the corresponding up and down states in the $Z$-basis. Notice that the states given by $\ket{\mc S_a,\,t=+1}$ in \eqref{eq: ferro scars} have exactly zero entanglement entropy, since they are simple product states of the form $\ket{\mc S_a,\,t=+1} = \otimes_i\ket{\psi_i}\,, ~\ket{\psi}_i = \cosh\left({\frac{\beta}{2}}\right)\ket{a}+\sinh\left({\frac{\beta}{2}}\right)\ket{a+1}$, where the sum $a+1$ is defined modulo 2. In contrast, the state $\ket{\bar{\mc S},\, t'=+1}$ obeys an area-law entanglement entropy, as it is generated by a finite-depth quantum circuit acting on a product state \cite{Eisert2010, Hermanns2017}.

\begin{figure*}
    \centering
    \subfloat[]{\includegraphics[width=0.482\linewidth]{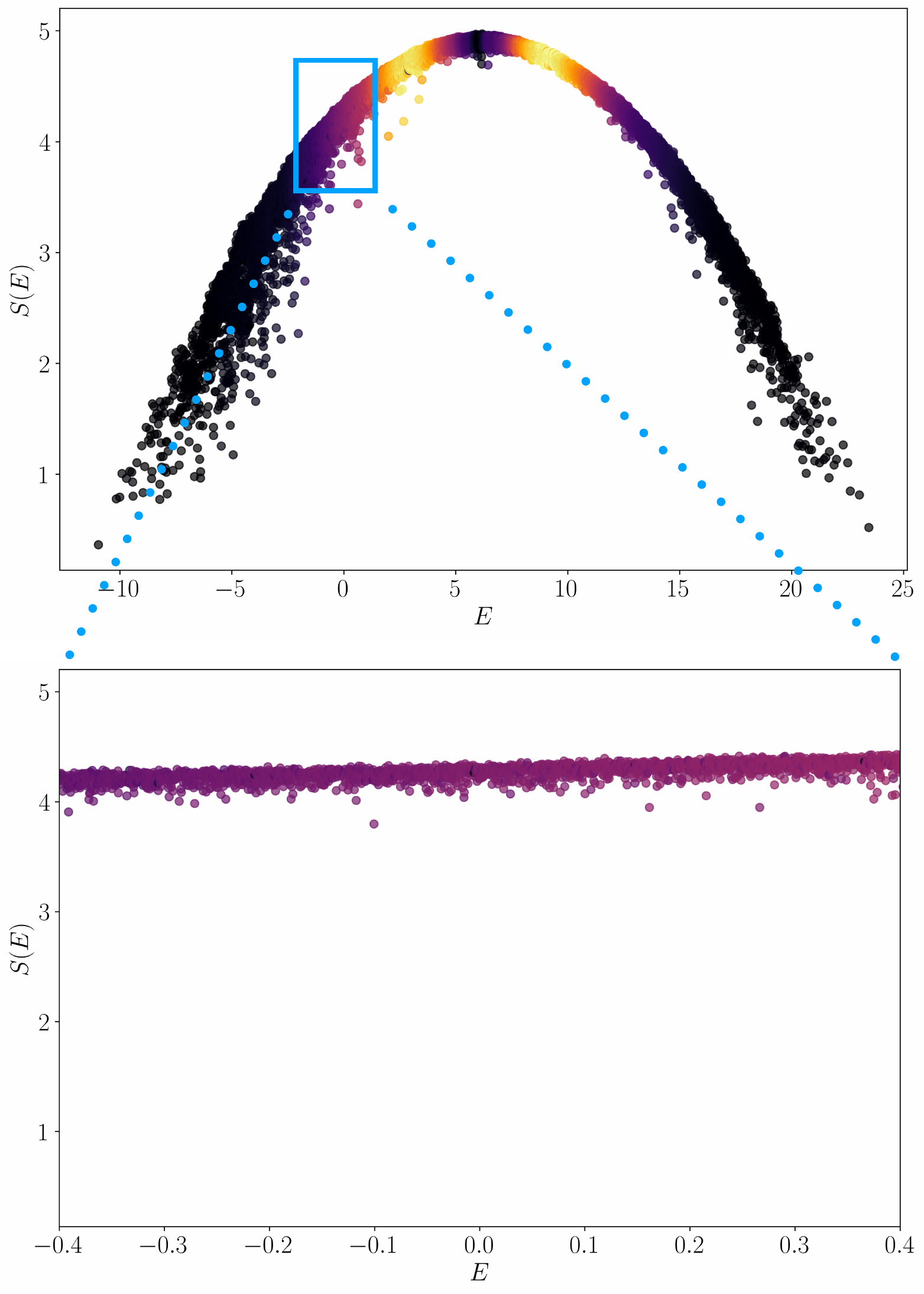}}
    \subfloat[]{
    \includegraphics[width=0.48\linewidth]{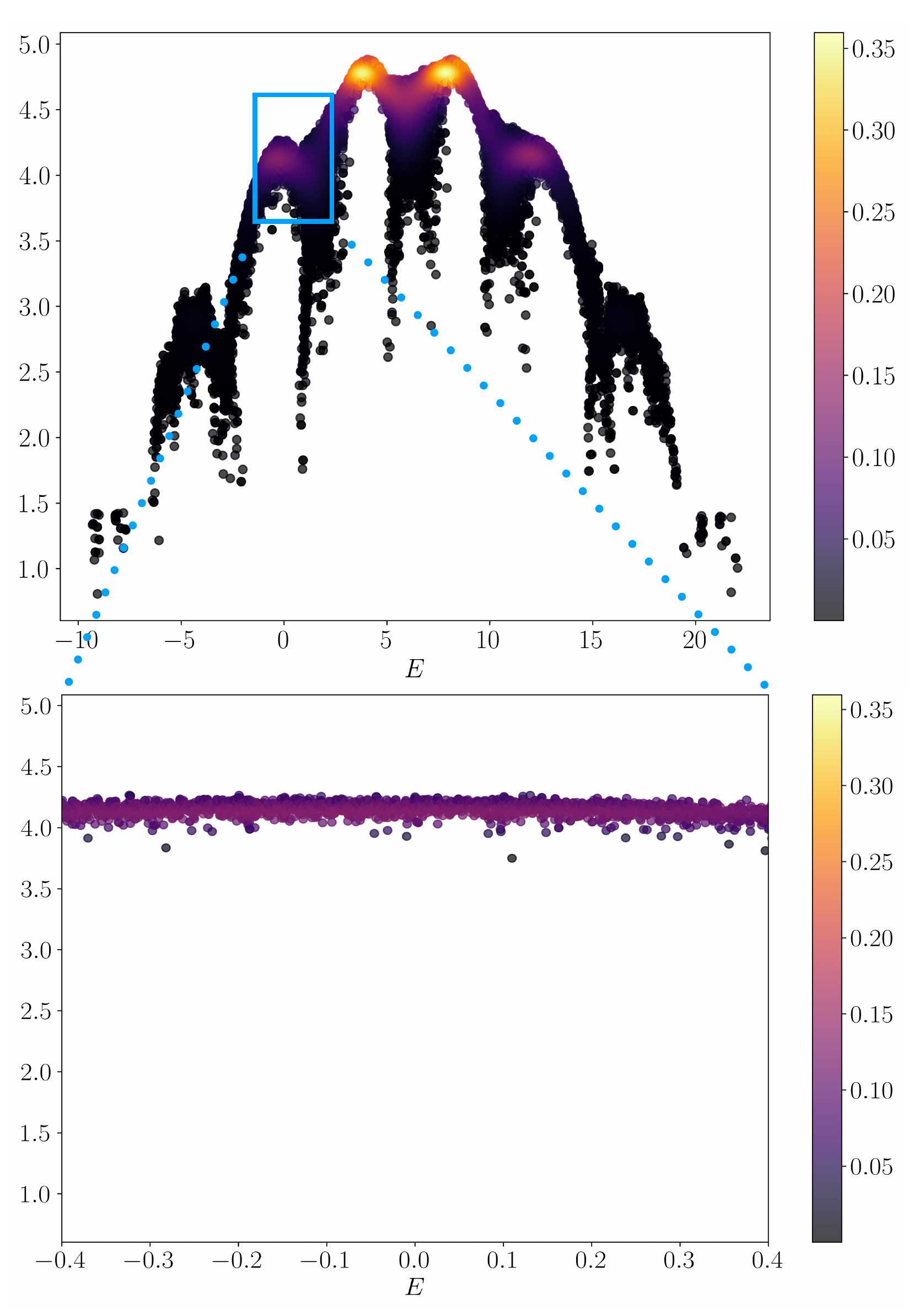}}
    \caption{\justifying Profile of entanglement entropy versus energy for (a) the twisted Hamiltonian $H_{\text{ND}}$ and (b) the twisted Hamiltonian $H_{\text{D}}$. Both plots are obtained from exact diagonalization of the corresponding Hamiltonians for a chain of size $L=16$. It is clear from the plot that there are no signs of low-entangled states at the $E=0$ line. Giving further evidence that the twisted sectors does not contain QMBS states at zero energy, for which the antisymmetric $\ket{\mc S_-}$ can be dual to.}
    \label{fig: twist_SvsE}
\end{figure*}

In the twisted sector, the Hamiltonian, though still a sum of projectors, admits no states annihilated by all terms: any $\ket{\phi}$ with $\mc Q^{(1)}\ket{\phi}=0$ necessarily has $\mc Q^{(-1)}\ket{\phi}\neq 0$, where $\mc Q^{(1)}$ and $\mc Q^{(-1)}$ are the untwisted and twisted projectors, respectively. This absence signals a breakdown of the SMF structure \cite{Castelnovo2005}. While the twisted Hamiltonian remains non-integrable, it lacks the structure needed to support non-thermal mid-spectrum states \cite{Chamon2019}, so we expect all states to obey the ETH conjecture. Indeed, a numerical analysis of entanglement profiles confirms no special states in the twisted sector, supporting the expectation that the twisted spectrum follows ETH. The result of the entanglement profile is displayed in Fig.\ref{fig: twist_SvsE}.

\section{Duality between QMBS subspaces \label{sec: Duality}} The full model $H$ admits a Kramers-Wannier duality that exchanges the coupling $J_1\Leftrightarrow J_2$. The duality can be understood in terms of a SQC\cite{Xie2023} $\mc U$, commonly written as
\begin{equation}
\mc U = \frac{1+\mathrm{i}Z_1Z_N}{\sqrt{2}}\prod_{j=N}^1 \frac{1+\mathrm{i}X_j}{\sqrt{2}}\frac{1+\mathrm{i}Z_jZ_{j+1}}{\sqrt{2}}\,.
\label{eq: sequential circuit}
\end{equation}
The above operator $\mc U$ is responsible for performing the KW duality, which performs $X_i\mapsto Z_iZ_{i+1}$ and $Z_{i-1}Z_{i}\mapsto X_i$ given that the global symmetry $\eta=\prod_i X_i=+1$, meaning that the KW duality only defines equivalences between symmetric states of the spectrum. One is able to generalize the duality, such that it is well defined for $\eta=-1$ states as well, at the cost of introducing symmetry twists $t$ \cite{Oshikawa2023,Vanhove2025}. The operator mapping is modified, to accommodate the twists, such that at the boundaries we have $X_N\mapsto t'Z_NZ_1$ and $tZ_NZ_1\mapsto X_1$ this has the effect of mixing charge and twists sectors, since under this modified KW duality, we have $\prod_i X_i\mapsto t'$ and $t\mapsto \prod_iX_i$. Therefore, we can identify that the sectors $(1,\,1)$ and $(-1,\,-1)$ are mapped into themselves while $(1,\,-1)$ and $(-1,\,1)$ are exchanged. 

This has an interesting consequence when we consider the effect of the duality over the QMBS states. First of all, consider the linear combinations of QMBS states on the degenerate side $\ket{\mc S_\pm} = \left(\ket{\mc S_0}\pm\ket{\mc S_1}\right)/\sqrt{2}$. Here we omit the twist symbol, since the scars only exist on the $(\eta,\,1)$ sector. Clearly, with regard to the global $\mathbb{Z}_2$ symmetry $\eta$, we have that $\eta\ket{\mc S_+}=+\ket{\mc S_+}$ and $\eta\ket{\mc S_-}=-\ket{\mc S_-}$. The SQC $\mc U$ establishes the following mapping between states
\begin{equation}
    \mc U \ket{\mc S_+} = \ket{\bar{\mc S}},\,\quad \mc U \ket{\bar{\mc S}} = \ket{\mc S_+}\,.\label{Dual Symmetric Scars}
\end{equation}
Meanwhile, for the $\ket{S_-}\in (-1,\,1)$ QMBS, we observe that it must have a corresponding state $\ket{\psi_{\mc T}}\in (1,\,-1)$ at exactly the same energy. As we argued earlier, the twisted Hilbert space can no longer support QMBS states, due to the lack of a SMF structure. Meaning that the corresponding state $\ket{\psi_{\mc T}}\in (1,\,-1)$ must be thermal-like. In Fig.\ref{fig: twist_SvsE} we show the entanglement profile of the Hamiltonians within the corresponding twisted sectors, giving further evidence of the lack of a dual area-law state at zero energy. 

\begin{figure*}
\subfloat[\label{fig: matrix_elements_ferro_antisymm}]{\includegraphics[width=0.327\linewidth]{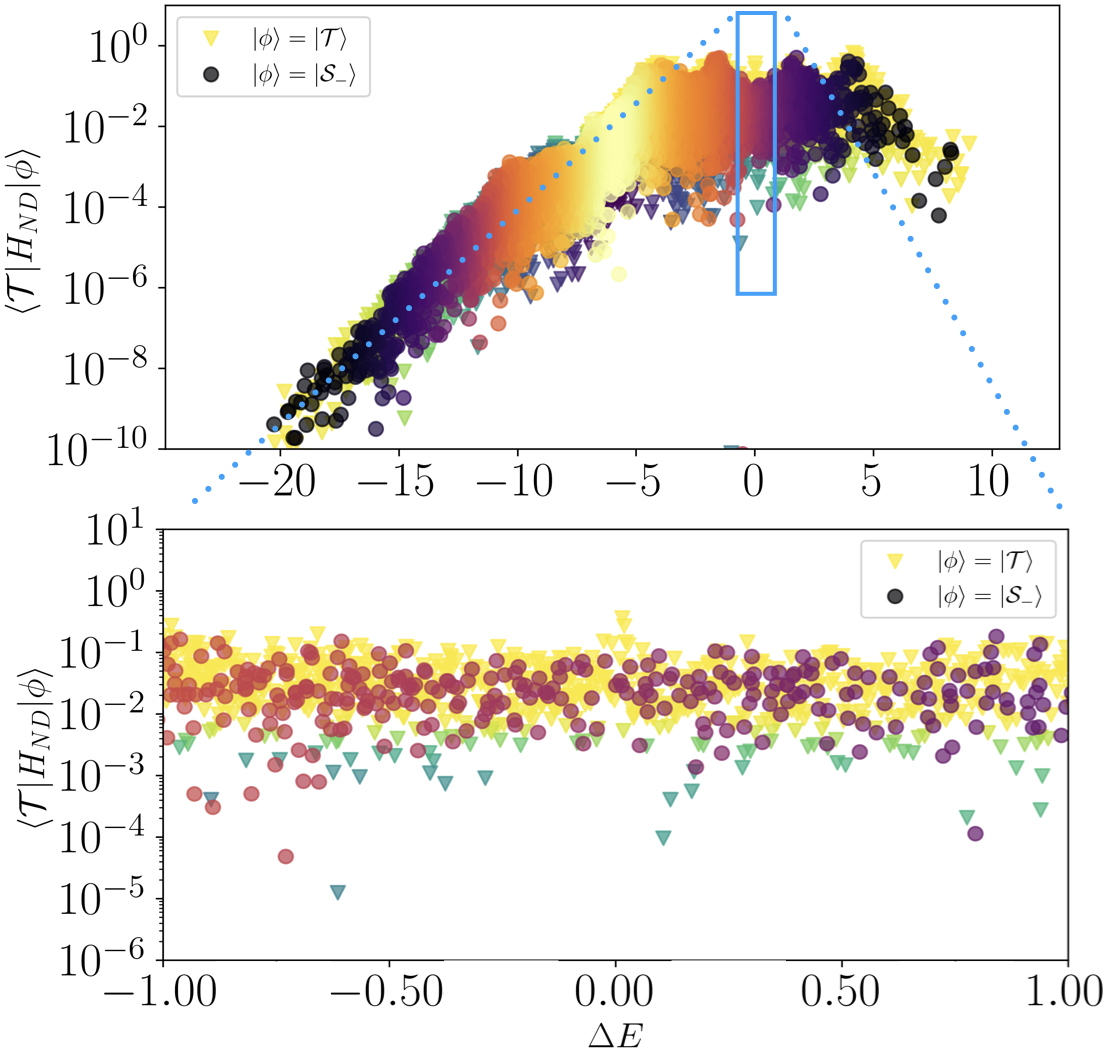}}
\hspace{1pt}
\subfloat[\label{fig: matrix_elements_ferro_symm}]{\includegraphics[width=0.327\linewidth]{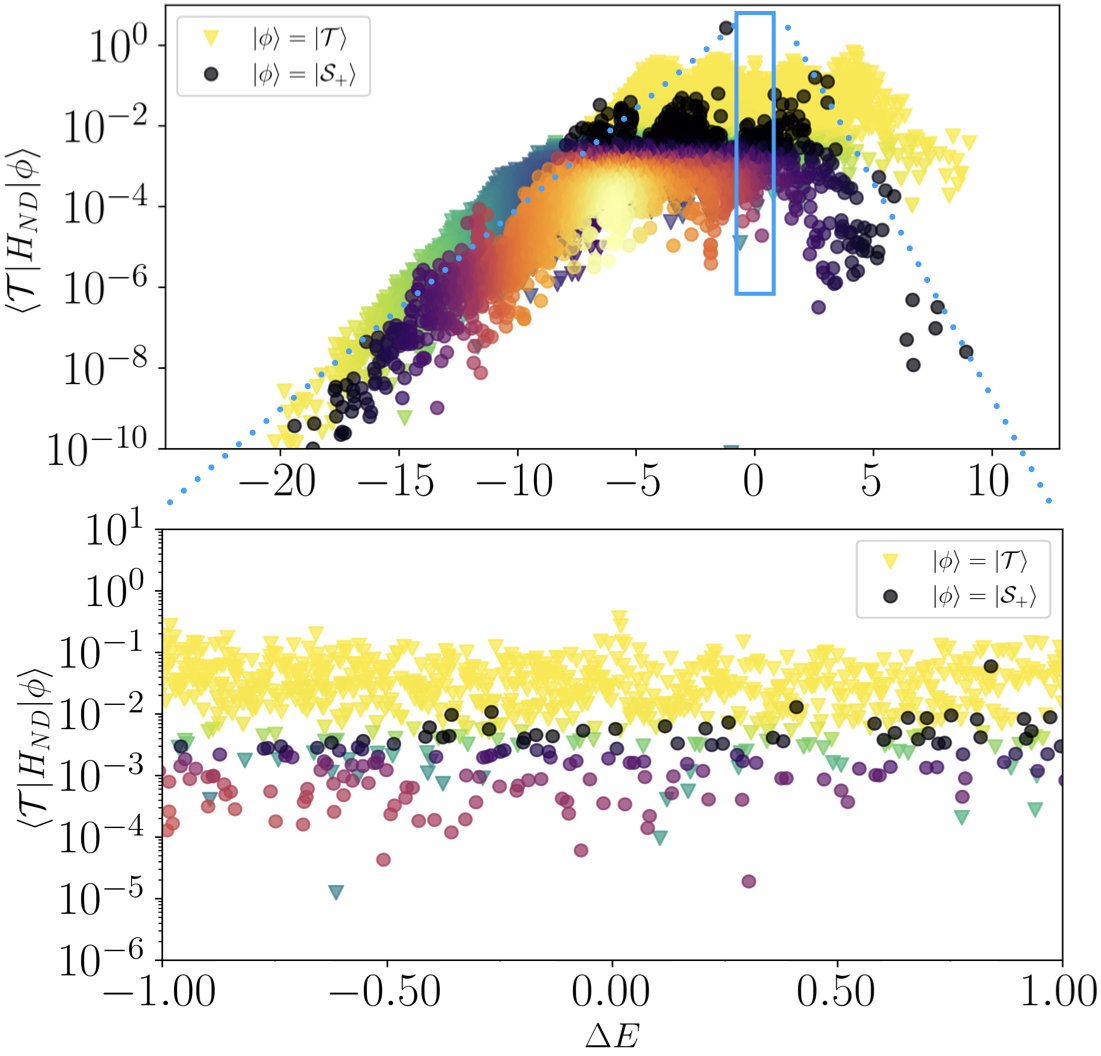}}
\hspace{1pt}
\subfloat[\label{fig: matrix_elements_para}]{\includegraphics[width=0.325\linewidth]{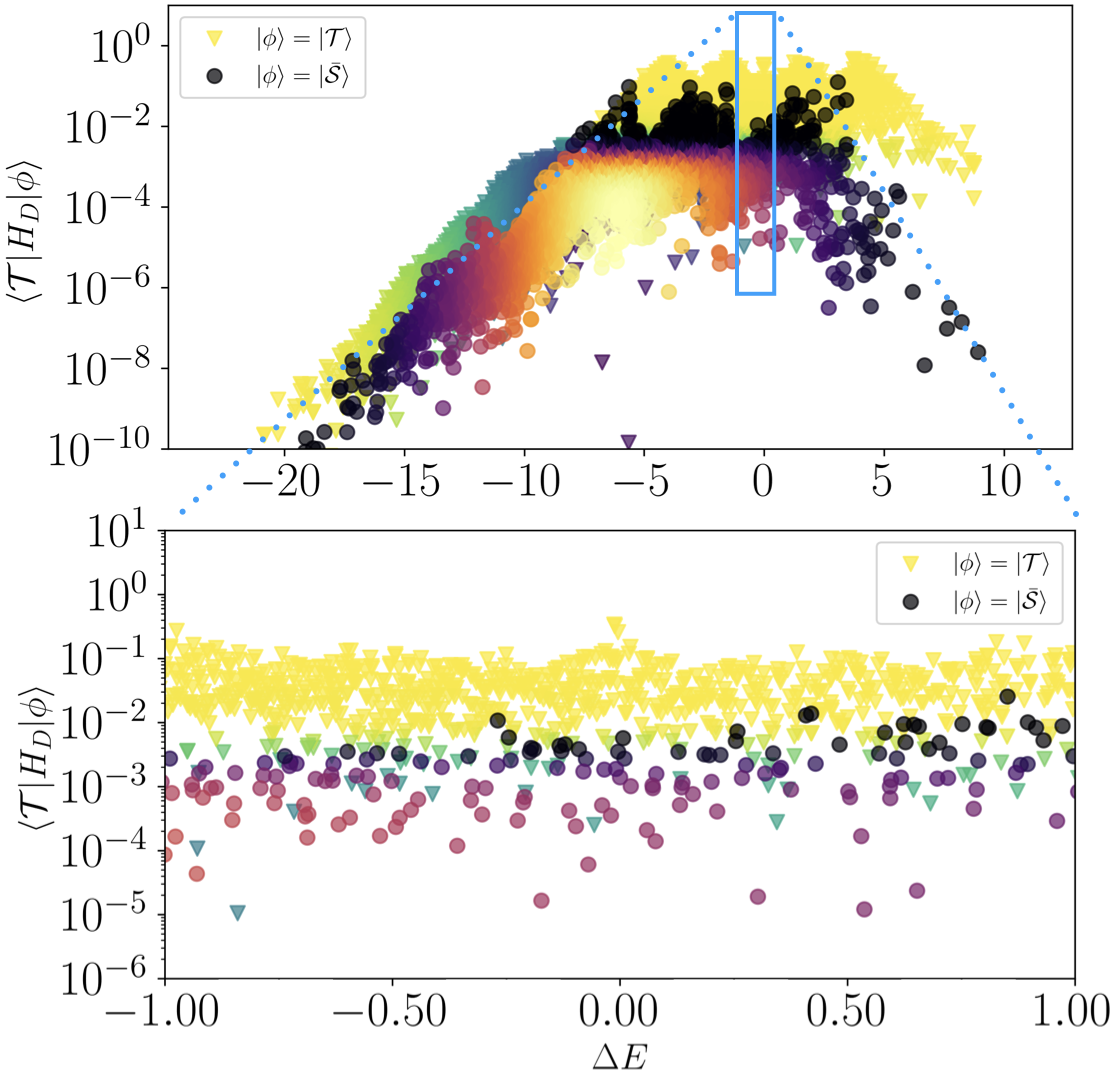}}
\caption{\justifying Matrix elements distributions of the corresponding perturbations $H_{\text{ND}}$ or $H_\text{D}$. The data was obtained from exact diagonalization for a chain of size $L=16$. The distribution for thermal states are represented by inverted triangles, meanwhile, the ones for scar states are represented by circles. We show the distributions for (a) the anti-symmetric scar $\ket{\mc S_-}$ in comparison to a thermal state. (b) the symmetric scar $\ket{\mc S_+}$ in comparison to a thermal state and (c) the symmetric scar $\ket{\bar{\mc S}}$ in comparison to a thermal state. Overall, the symmetric ones have a clear distinct behavior, with a suppression near $\Delta E\approx 0$, meanwhile the anti-symmetric scar have a thermal-like distribution.}
\label{fig:matrix elements}
\end{figure*}
\section{Signatures of the QMBS states in presence of the dual Hamiltonian \label{sec: scars signatures}} At this point, we established a relation between the QMBS states (\ref{Dual Symmetric Scars}) and argued that the antisymmetric scar $\ket{\mc S_-}$ is mapped into a thermal-like state. To understand their differences, we now study the distribution of their matrix elements. For clarity, let us consider that we start at the regime with $J_2=0$, where the exact QMBS states are $\ket{\mc S_\pm}$, and we compute the matrix elements of $H_{\text{ND}}$ between the exact QMBS and all the other states in the spectrum. According to the duality relation, we can relate the matrix element distribution of these states as follows
\begin{align}
    \braket{\mc T|H_{\text{ND}}|S_+} = \braket{ \Tilde{\mc T}|H_{\text{D}}|\bar{\mc S}}\,, \nonumber~\braket{\mc T|H_{\text{ND}}|\mc S_-} = \braket{\Tilde{\mc T}|H_{\text{D}}|\psi_{\mc T}}\,,
\end{align}
where $\ket{\Tilde{\mc T}}$ represents some dual thermal-like state. The above relation tell us that, if we observe some violation of ETH on the matrix element distribution of $\ket{\mc S_+}$ the same should happen for $\ket{\bar{\mc S}}$, meanwhile, $\ket{\mc S_-}$, is expected to have a more fragile behavior as we turn on the dual part $H_{\text{ND}}$, meaning it should present a stronger hybridization with other nearby states. We show the results in Fig.\ref{fig:matrix elements} for a chain of size $N=16$, obtained through exact diagonalization. As expected from the analytical results, the antisymmetric QMBS state has a distribution similar to that of a thermal state, as displayed in Fig.(\ref{fig: matrix_elements_ferro_antisymm}), signaling that this state should be much more prone to hybridization as soon as the perturbation is increased.

Meanwhile, for the symmetric scar states $\ket{\mc S_+}$ and $\ket{\bar{\mc S}}$, we can observe a significant difference on the behavior of matrix elements in comparison to the distribution of thermal states, which is displayed in Fig.\ref{fig: matrix_elements_ferro_symm} and Fig.\ref{fig: matrix_elements_para}, respectively. Away from the region of vanishing energy gap $\Delta E= E_{\mc T}-E_{\mc S}\approx 0$ the distributions behaves very similar, decaying exponentially fast, as expected from random matrix theory (RMT) \cite{DAlessio2016}. The more significant difference happens at the region $\Delta E\approx 0$, where we see that the matrix elements with the scar states are more suppressed in comparison to the thermal ones, which explicit their non-thermal behavior. 

\begin{figure*}
    \centering
    \subfloat[\label{Fig:Ferro_elements_L12vsL16}]{\includegraphics[width=0.49\linewidth]{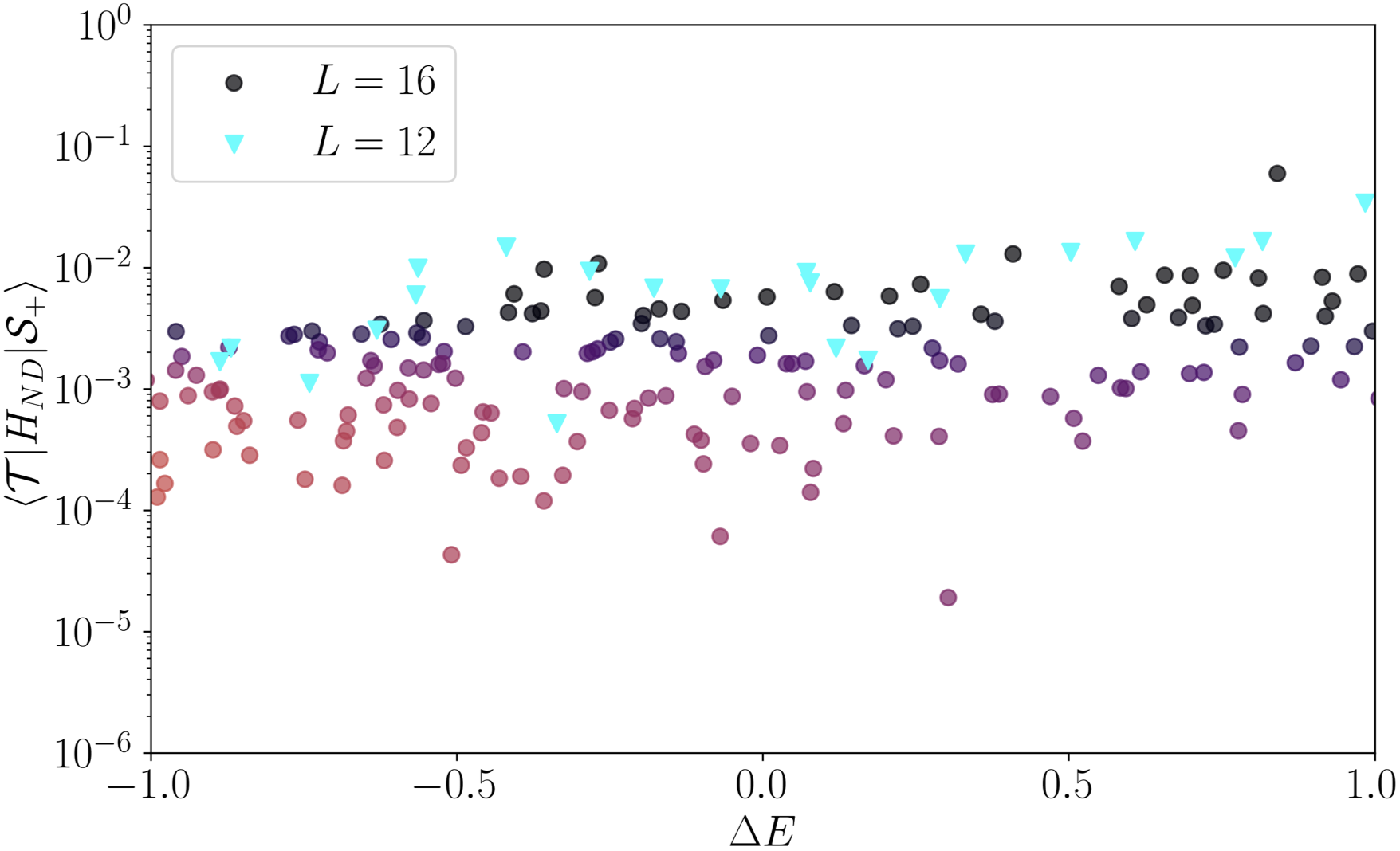}}
    \hspace{1pt}
    \subfloat[\label{Fig: Para_elements_L12vsL16}]{\includegraphics[width=0.49\linewidth]{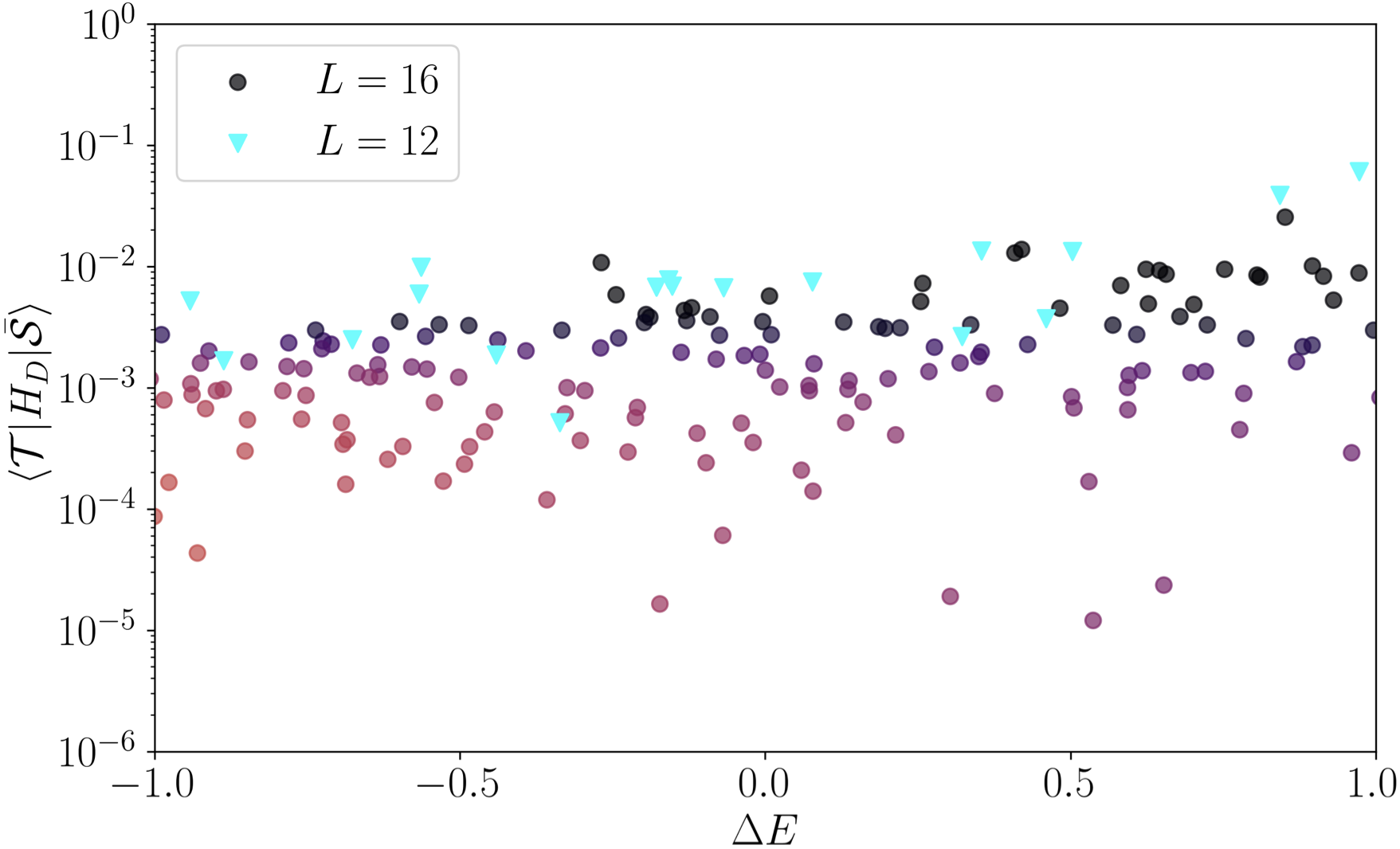}}
    \caption{\justifying Distribution of the matrix elements for both (a) $\ket{\mc S_+}$ and (b) $\ket{\bar{\mc S}}$ QMBS states in an energy window around the $\Delta E=0$ point, for sizes $L=12$ (cyan) and $L=16$. The suppression for the matrix elements involving the scar states, does not seem to decrease significantly with system size.}
    \label{fig:system_size_comparison}
\end{figure*}

It is important to mention that although this analysis makes evident the difference between the QMBS states we analyze, this is not enough to ensure that these states are in fact robust in the current scenario. As pointed out in \cite{Motrunich2020}, this suppression of the off-diagonal matrix elements involving the scar states may not be enough to prevent thermalization in the thermodynamic limit. For instance, consider the plot in Fig.\ref{fig:system_size_comparison}, where we compare the distributions for sizes $L=12$ and $L=16$. Even for smaller sizes, the matrix elements involving the scar states are roughly of the same order. Which suggests that, indeed, as one takes the thermodynamic limit, thermalization will unavoidably happen due to the fact that the matrix elements might not decrease fast enough to compensate for the closure of the energy gaps. Similar observations have been made in the analysis of the PXP model in the presence of some particular perturbations \cite{Motrunich2020}. Instead, what this analysis suggests is that these states can experience a slow thermalization phenomena in comparison to the typical thermal states, with a thermalization time $t(\lambda)\sim \lambda^{-1/2}$, with $\lambda$ the appropriate perturbative parameter, as predicted by the authors of \cite{Motrunich2020}.

\section{Fidelity loss of QMBS states \label{sec: fidelity loss}}
\begin{figure*}
    \subfloat[]{\includegraphics[width=0.488\linewidth]{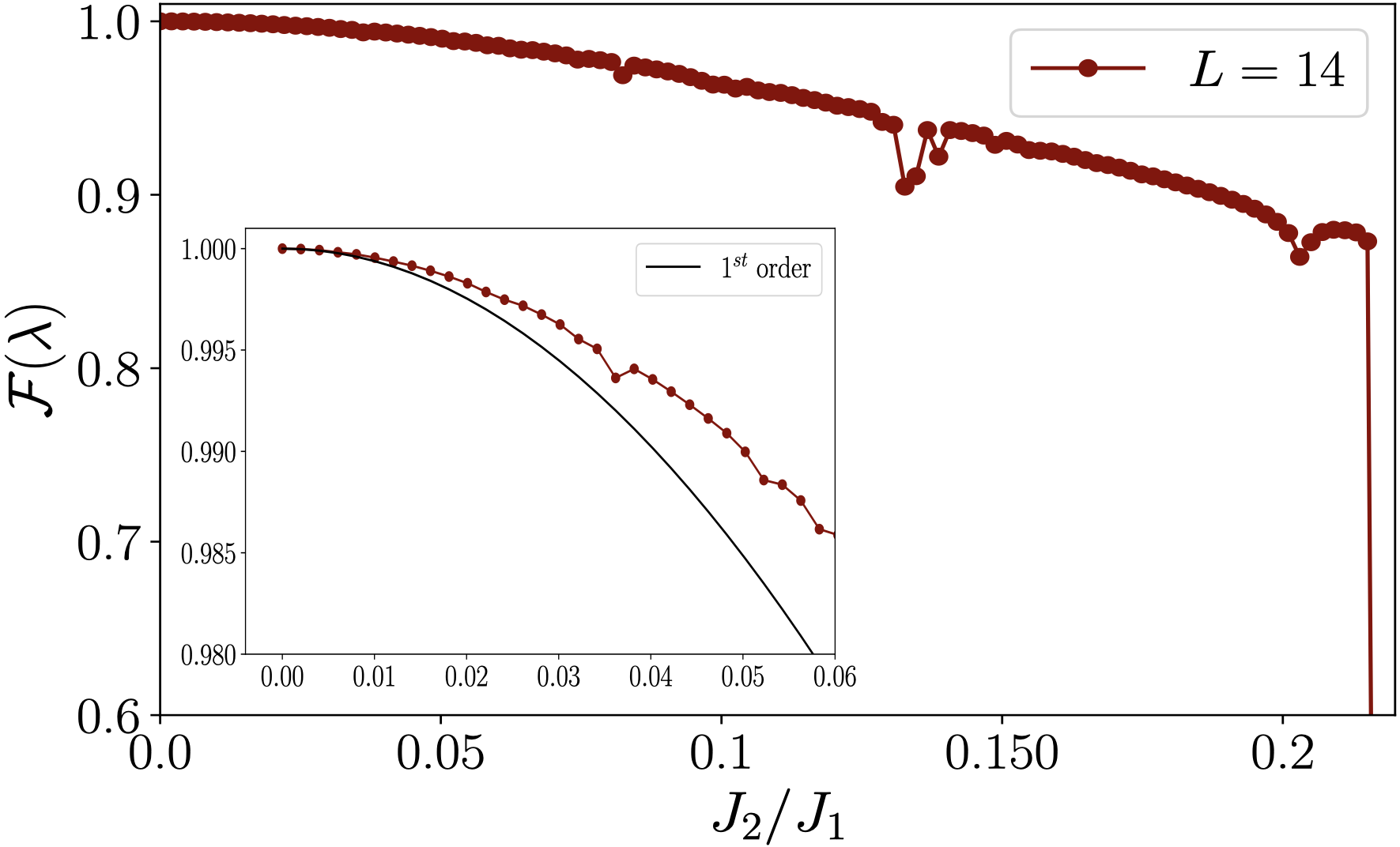}}
    \subfloat[\label{fig: Fid_perturbation_nondeg_to_deg}]{
    \includegraphics[width=0.488\linewidth]{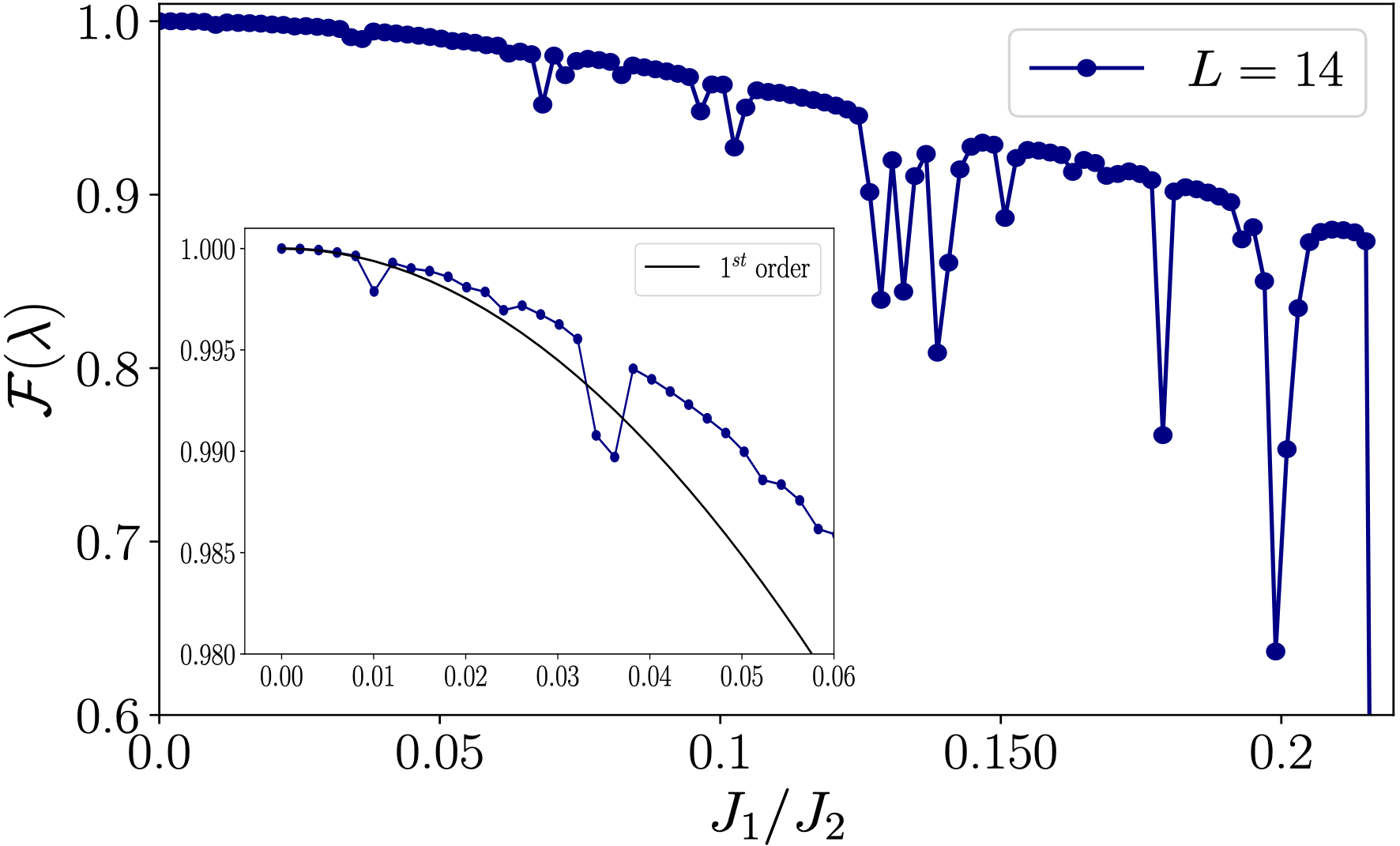}}
    \caption{\justifying Comparison of the ED numerical fidelities (red and blue) for the symmetric scar states $\ket{\mc S_+}$ and $\ket{\bar{\mc S}}$, respectively. The inset in each figure display the comparison with the theoretical first order (black solid line) approximation. (a) when $H_{\text{ND}}$ is taken as the perturbation term and (b) when $H_{\text{D}}$ is taken as the perturbation term.}
    \label{Fig: Fid_loss_theory}
\end{figure*}

Regarding the stability of the symmetric scar states, we can further study their fidelity loss as we interpolate between the two subspaces. Once again, consider the full model $H(\lambda)=H_{\text{D}}+\lambda H_{\text{ND}}$, where $\lambda=J_2/J_1$, with the fidelity given by  $ \mc F(\lambda) = \left|\braket{\mc S(\lambda=0)| \mc S(\lambda)}\right|^2$. The state $\ket{\mc S(\lambda)}$ represents the corresponding scar obtained after diagonalizing $H(\lambda)$. The behavior of the fidelity for the QMBS $\ket{\mc S_+}$ as $\lambda$ is increased is shown in Fig.\ref{Fig: Fid_loss_theory}. We see that, at least for a certain range of $\lambda$, the perturbed QMBS state remains at high fidelity, signaling some robustness (at least for finite size systems) even in the presence of the additional Hamiltonian perturbation. 

The same analysis can be performed for the fidelity loss of the scar $\ket{\bar{\mc S}}$. Consider that we take the inverse route with the Hamiltonian $H=H_{\text{ND}}+\frac{1}{\lambda}H_{\text{D}}$, and we now compute the fidelity of the exact scar $\ket{\bar{\mc S}}$ as we increase the value of $1/\lambda$, the result is shown in Fig.\ref{Fig: Fid_loss_theory}. In this situation, we once again see a range of high fidelity, although we have the presence of an increased number of sudden drops in the fidelity, which can be attributed to accidental hybridization \cite{Motrunich2020}. Essentially, when eigenstates are very close in energy, their weights can spread over nearby states, resulting in a lower overlap between them. This accidental hybridization is a consequence of avoided level crossings that can occur in the middle of the spectrum, but when unavoidably the energy levels crosses, the overlap recovers, as pointed out in \cite{Motrunich2020}. We see that, as we increase the value of the coupling $\lambda$, the fidelity between the states decreases, but in a very slow way. Signaling that although these scars may not be robust \cite{Motrunich2020}, the state can still be found for some small range of parameters.

As a consequence of the suppression observed in the off-diagonal matrix elements for the symmetric QMBS states \ref{sec: scars signatures}, we could employ a perturbative analysis to understand some overall behavior of these states. We now proceed to show that, in fact, first order perturbation theory gives a very good approximation - at least in this finite size analysis - and the error we commit is, essentially, negligible. For that, we consider the approximation to the perturbed states
\begin{equation}
    \ket{\mc S(\lambda)} = \ket{\mc S} + \lambda \sum_{n:\, E_n\neq E_{\mc S}}\frac{\braket{n^{(0)}|V|\mc S}}{E_n^{(0)} - E_{\mc S}^{(0)}}\ket{n^{(0)}}\,,
    \label{eq: pert_scars}
\end{equation}
where $\ket{\mc S}$ corresponds to an unperturbed scar state and $V$ represents a local perturbation (in our case, either $H_{\text{ND}}$ or $H_\text{D}$). Clearly, the validity of such approximation is tied to the condition that the elements $\left|\frac{\braket{n^{(0)}|V|\mc S}}{E_n^{(0)} - E_{\mc S}^{(0)}}\right|$ are finite and, at most, of $\mc O(1)$. Therefore, in principle, one should not expect any agreement with the numerical results and such a naive approximation. Any finite matrix element between the scar states and the neighboring thermal ones should give us a diverging contribution, since the energy gap $E_n^{(0)}-E^{(0)}_{\mc S}$ is exponentially small between the states, as predicted by random matrix theory (RMT) \cite{DAlessio2016}, where the overall behavior of matrix elements should abide by the ETH conjecture
\begin{equation}
    \mc O_{mn}\sim \bar{\mc O}\delta_{mn}+\sqrt{\frac{\bar{O^2}}{\mc D}}R_{mn}\,,
    \label{eq: ETH_matrix elements}
\end{equation}
where $\mc D$ is the dimension of the full Hilbert space, $R_{mn}$ is a random variable and $\bar{O}$ is just a simple average of the diagonal elements, i.e., $\frac{1}{\mc D}\sum_i O_{ii}$. Eq.\eqref{eq: ETH_matrix elements}, predicts that matrix elements between two different states are exponentially small and goes to zero as a power of $\mc D^{-1/2}$. 

On the other hand, we expect that the distance between two different energy levels should scale as $1/\mc{D}$, which results in divergence in \eqref{eq: pert_scars} of the order $\mc D^{1/2}$, spoiling the perturbative analysis. In order to have a glimpse of how good are the results drawn from a perturbative analysis, we compare the results obtained from it to understand the fidelity loss of our states. The fidelity loss of a given state, with a perturbative correction \cite{Gu2010} up to first order can be simply written as
\begin{align}
    \mc F_P(\lambda) &= \frac{|\braket{\mc S(0)|\mc S(\lambda)}|^2}{\braket{\mc S(\lambda)|\mc S(\lambda)}}\nonumber\,\\
    &=\left[{1+\lambda^2\sum_{n\neq\mc S}\frac{|\braket{n^{(0)}|V|\mc S(0)}|^2}{(E^{(0)}_n - E^{(0)}_{\mc S})^2}}\right]^{-1}\,.
\end{align}
From inspecting Fig. \ref{Fig: Fid_loss_theory}, up to the sudden jumps in the ED results, one can see that the perturbative analysis is able to capture the overall behavior of fidelity loss with a good precision. With the overall values, at this system size, differing by roughly 0.4\%. As the perturbative parameter $\lambda$ is increased, the matching becomes worse and we exit the perturbation valid region. The fact that we can get such a nice agreement between the numerical data and the theoretical curve is quite remarkable, and can be attributed to the suppression between the matrix elements involving the scar states. With the increase in system size, the range in which perturbation theory may give sensible results may start to reduce, due to the increase in the density of level crossings and to the fact that the matrix elements may not decrease fast enough to compensate the divergence coming from the closure of the energy gap, which would spoil the perturbative analysis. It is also worth to stress, that although one may be tempted to consider higher order corrections in the perturbative series, this does not help. Curiously, it makes the predictions worse. This is due to the fact that at any order higher than the first, the perturbative series will include terms like $\frac{\braket{\mc T|V|\mc T}}{E_{\mc S}-E_{\mc T}}$, which are unavoidably divergent. 

Therefore, surprisingly, the nice agreement of numerical results and perturbation theory is restricted to first order. Finally, we also present the plot of the entanglement entropy versus energy  for different and fixed values of the $\lambda$ coupling as we interpolate from the $H_{\text{D}}$ Hamiltonian to the $H_{\text{ND}}$. For completeness, in the appendix \ref{app: Fidelity} we provide more details on the computation of the fidelity using exact eigenstates for finite system sizes. 

\section{Conclusions and Outlook \label{conclusions}} In this work, we examined a simple model exhibiting exact QMBS states at different fixed points and explored how they are related through a KW duality. We showed, both analytically and numerically, that the duality relation can be used to infer whether a QMBS state preserves its non-thermal properties, depending solely on whether the dual state is thermal-like or non-thermal. 

This suggests that the additional structure introduced by the duality can shed light on the different degrees of robustness that a QMBS state may exhibit. In this sense, it provides a finer level of characterization beyond the current state-of-the-art understanding established in Ref. \cite{Motrunich2020}. In particular, it indicates that thermalization times for individual scar states may be further enhanced when such additional structures are present. More broadly, this points to a new avenue for investigating the robustness or fragility of QMBS, by exploiting dualities and their action across twisted symmetry sectors. We note that dualities have recently been proposed as a mechanism for generating new scar states in 2+1 dimensions \cite{Katsura2025}, although the broader implications for robustness were not the focus of that work.

We expect this behavior to emerge in a wide range of settings, particularly in higher dimensions, enabling the identification of new quantum many-body scar (QMBS) states and distinguishing those that are genuinely non-thermal. For generic non-invertible dualities, such as Kramers–Wannier, a subset of QMBS typically thermalizes due to additional structures (e.g., twist sectors) required to track state mappings—structures that disrupt the conditions for scarring. By contrast, invertible dualities, which require no such structures, are expected to stabilize all scarred states. This distinction establishes duality as a robust criterion for diagnosing and predicting non-ergodic behavior beyond fine-tuned models. Moreover, with digital quantum computers already used to explore topological order\cite{satzinger2021realizing} and correlated dynamics\cite{cochran2025visualizing} via sequential circuits, our Hamiltonian and duality protocol are well-suited for near-term experimental realization.
\section{Acknowledgments} We are thankful to Hernán B. Xavier for enlightening conversations during the completion of this work and to Hosho Katsura for carefully reading the manuscript and making insightful observations. ED calculations were performed with the aid of QuSpin package \cite{Quspin1, Quspin2} in Python. This research has been supported by the MOST Young Scholar Fellowship (Grants No. 112-2636-M-007-008- No. 113-2636-M-007-002- and No. 114-2636-M-007 -001 -), National Center for Theoretical Sciences (Grants No. 113-2124-M-002-003-) from the Ministry of Science and Technology (MOST), Taiwan, and the Yushan Young Scholar Program (as Administrative Support Grant Fellow) from the Ministry of Education, Taiwan.

\bibliography{StabScarsref}
\bibliographystyle{my_apsrev4-2}

\appendix

\begin{figure*}
    \centering
    \includegraphics[width=0.5\linewidth, scale=0.25]{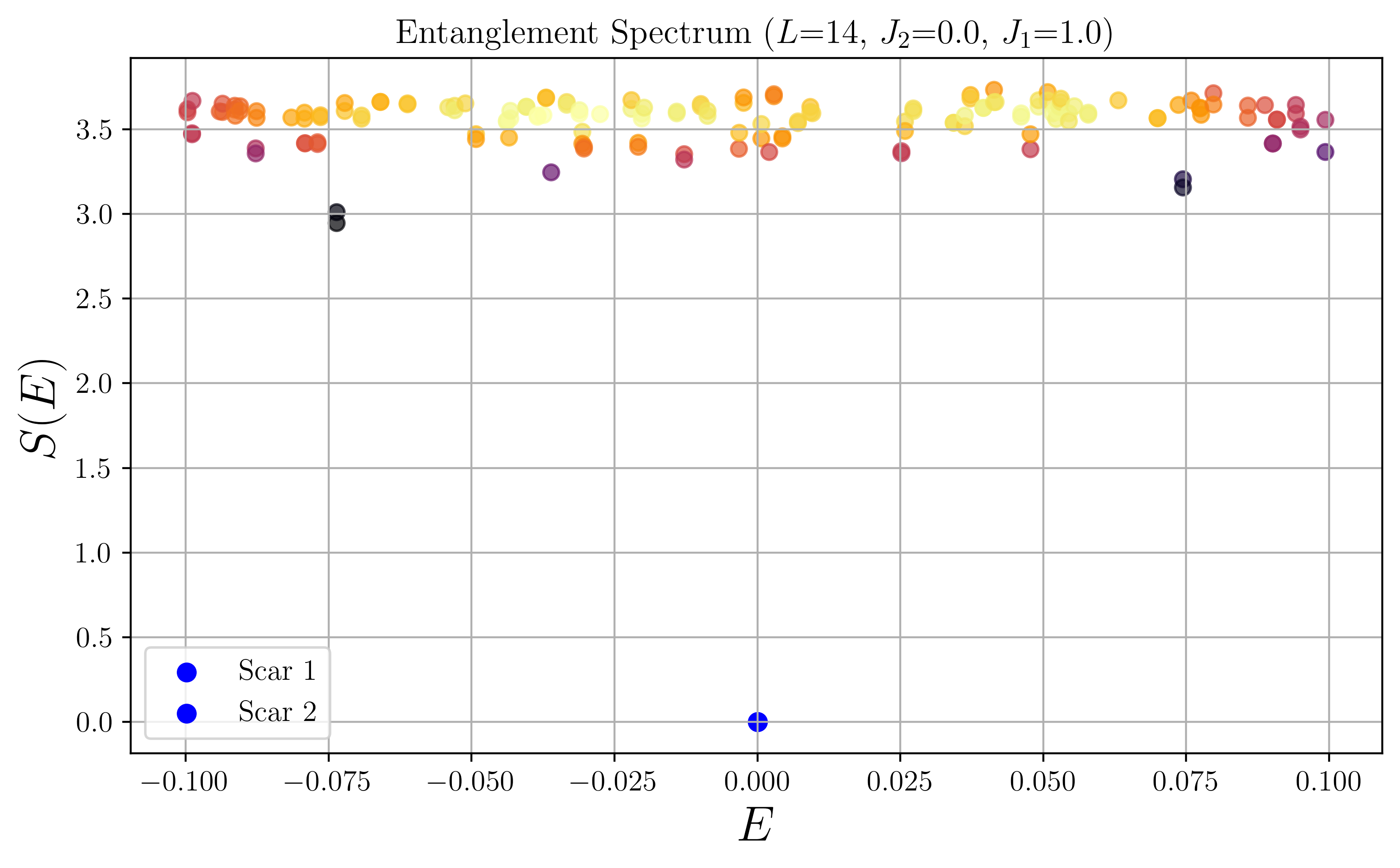}\includegraphics[width=0.5\linewidth, scale=0.25]{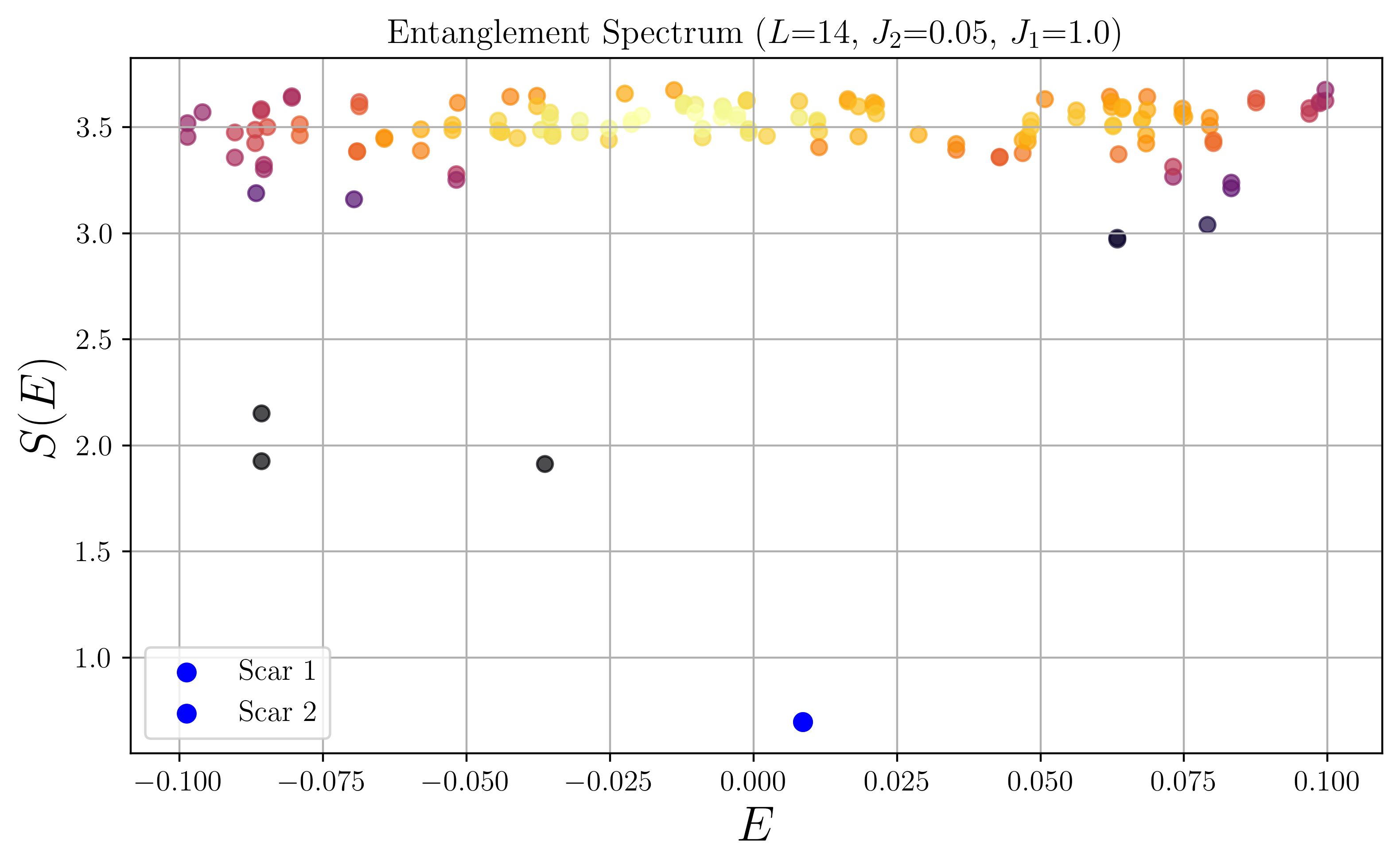}  \includegraphics[width=0.5\linewidth, scale=0.25]{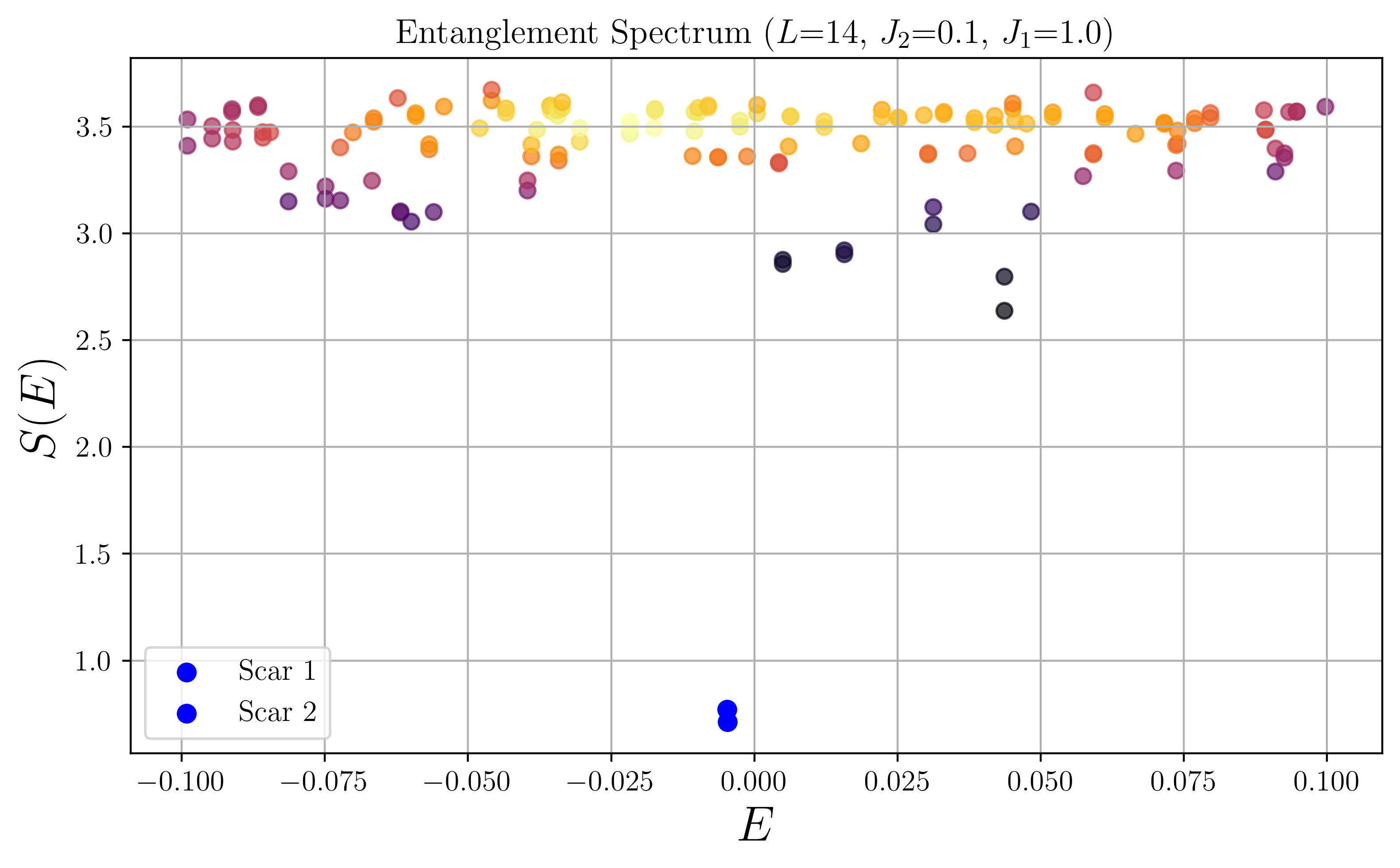}\includegraphics[width=0.5\linewidth,scale=0.25]{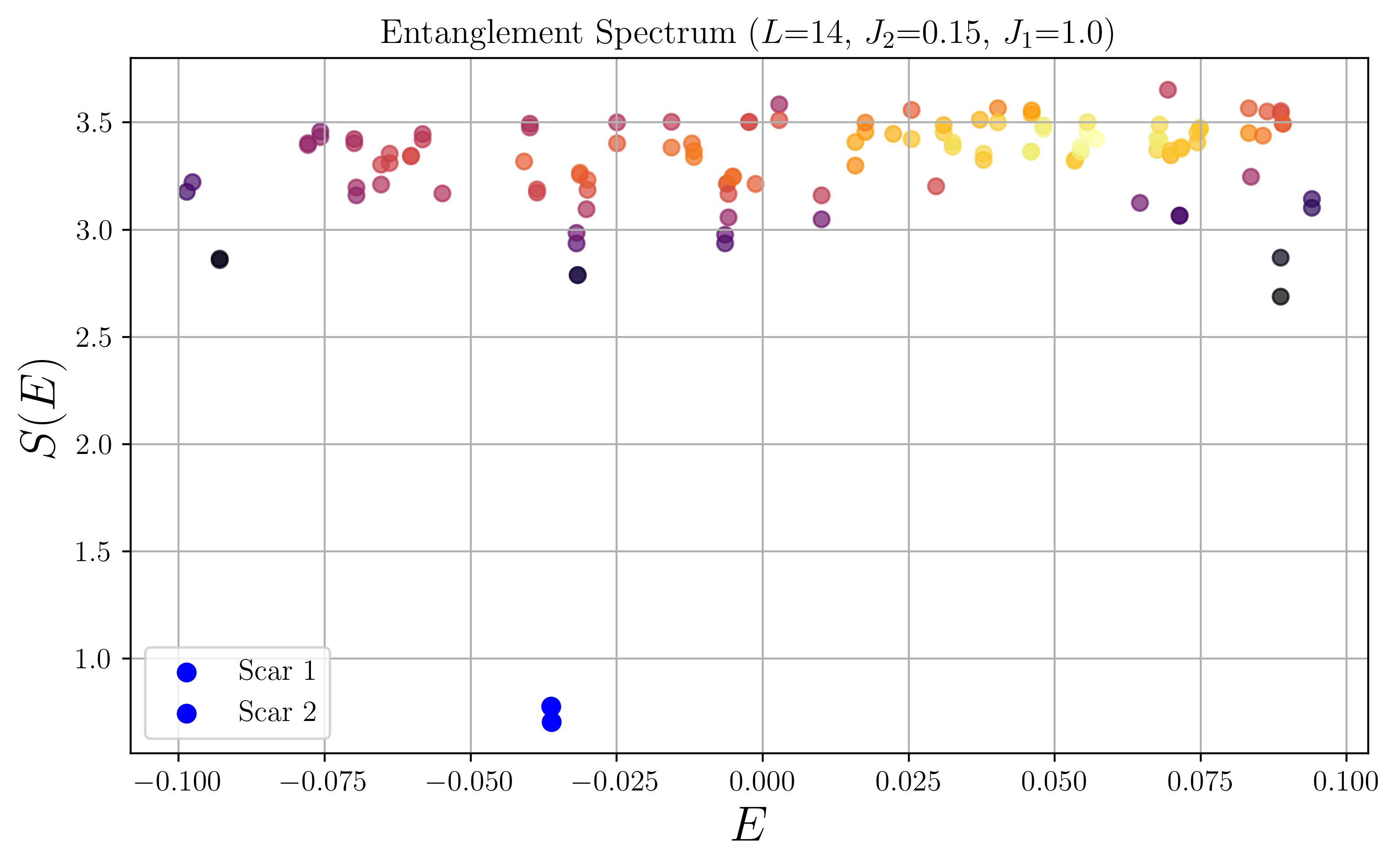}  
    \includegraphics[width=0.5\linewidth, scale=0.25]{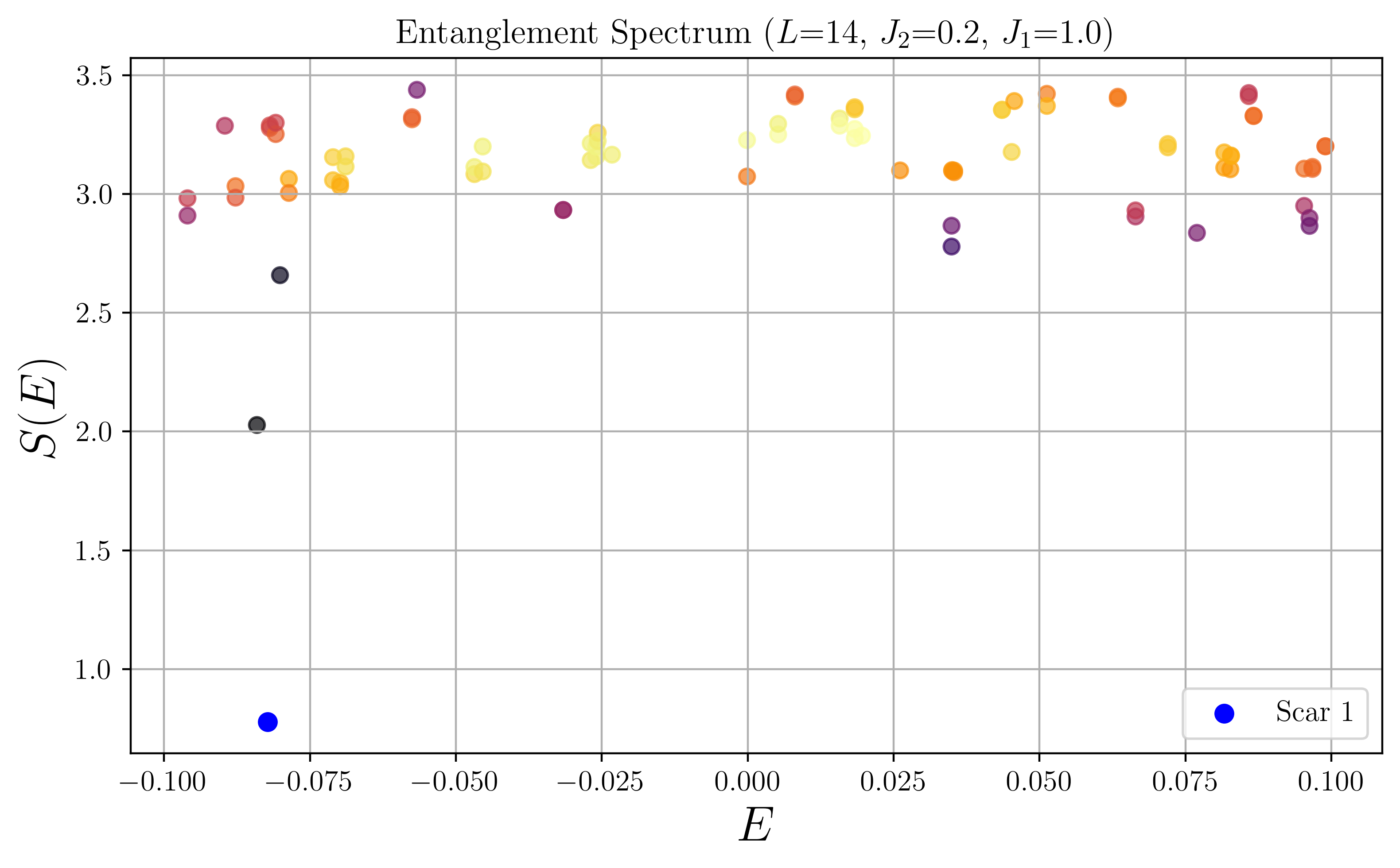}\includegraphics[width=0.5\linewidth, scale=0.25]{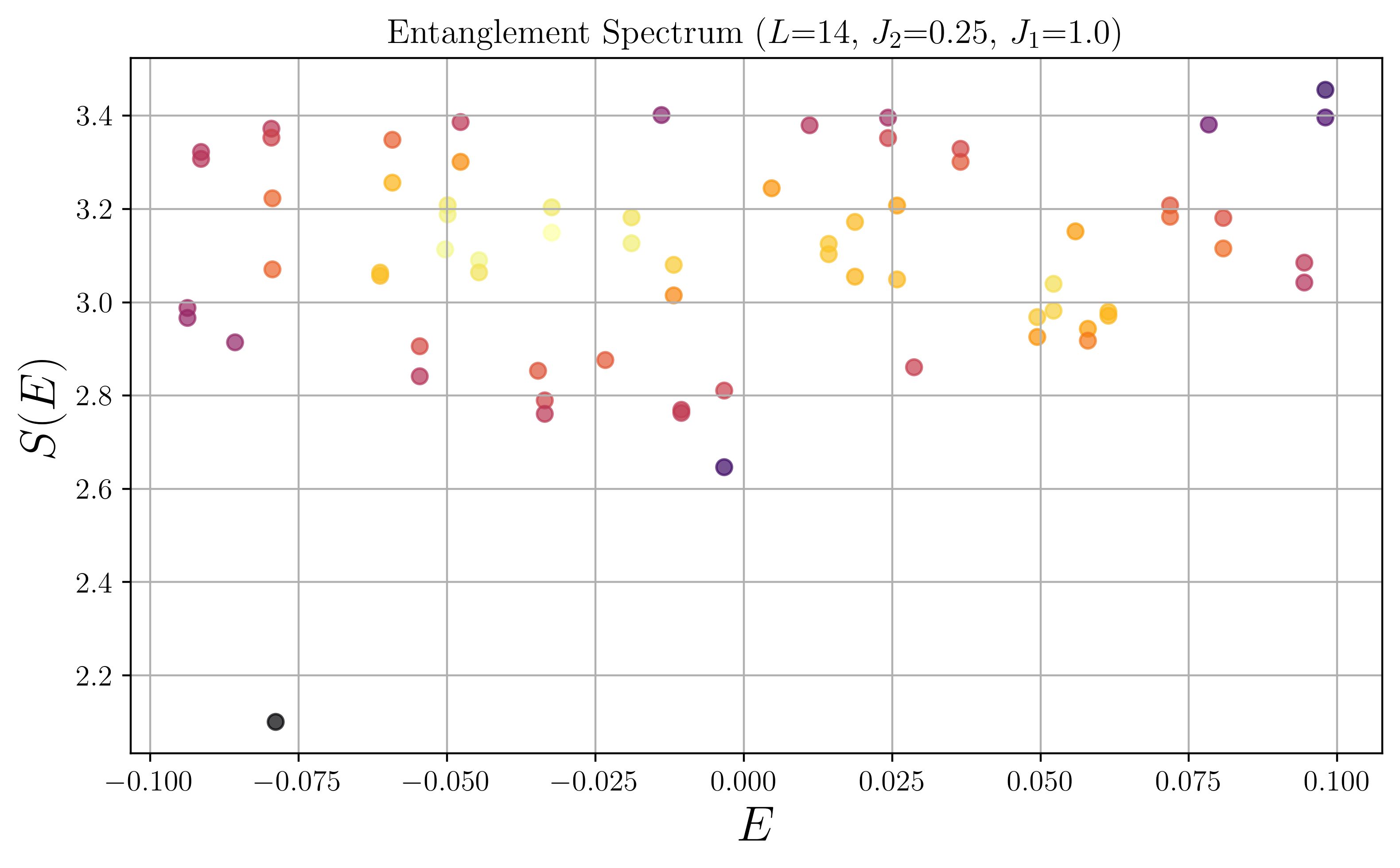}
    \caption{\justifying Entanglement entropy versus energy obtained from an exact diagonalization procedure of a chain of size $L=14$. }
    \label{fig:entanglement-fidelity}
\end{figure*}
\section{Further details of the QMBS Hamiltonians \label{app: A}}
\subsection{Non-degenerate QMBS subspace}
First, let us analyze each term of the Hamiltonian $H=H_{\text{D}}+\lambda H_{\text{ND}}$ individually. Consider the following model defined on a one-dimensional lattice $\Lambda$ of $N$ sites 
\begin{align}
     H_{\text{ND}}(\beta)&=\sum_i \alpha_i\left(e^{-\beta\,Z_i\left(Z_{i-1}+Z_{i+1}\right)}-X_i\right)\,,
    \label{eq: Hamiltonian para scar}
\end{align}
where $\alpha_i$ generically is a site-dependent coupling and $X,\,Z$ stand for the corresponding Pauli matrices. This Hamiltonian belongs to a class of models known as stochastic matrix form (SMF) decomposable Hamiltonian's \cite{Castelnovo2005}. Notice that the terms in the Hamiltonian can be written as $H_{\text{ND}}=\sum_i \alpha_i \mathcal{Q}_i$, with $\mc{Q}_i$ a positive semi-definite operator ($\mathcal{Q}_i^2=2\cosh{(\beta Z_iZ_{i+1})} \mathcal{Q}_i$).

The Hamiltonian admits zero energy states $\ket{\bar{S}(\beta)}$, that annihilated by each of the $\mc Q_i$ operators. Whether $\ket{\bar{\mc S}}$ corresponds to the ground state or a scar state, depends solely on the coupling $\alpha_i$. If $\alpha_i$ is strictly positive on every site of the lattice, $\ket{\bar{\mc S}(\beta)}$ will describe the ground state physics of the Hamiltonian \eqref{eq: Hamiltonian para scar}, if $\alpha_i$ is allowed to admit negative values, then $\ket{\bar{\mc S}(\beta)}$ is an isolated zero energy scar state sitting in the middle of a dense spectrum, as it can be seen from the Fig.(\ref{fig:entanglement-para-scar+level-spacing}). Furthermore, the $H_{\text{ND}}$ is generically non-integrable as suggested by its level spacing statistics in Fig.\ref{fig:entanglement-para-scar+level-spacing}. 

The $\ket{\bar{\mc S}(\beta)}$ state admits other representations that are useful to compute correlations and to understand its properties. In particular, it can be written as a tensor network state
\begin{equation}
    \ket{\bar{\mc S}(\beta)} = \sum_{\left\{\sigma=0,1\right\}}\text{Tr}\left(A^{\sigma_1}\dots A^{\sigma_N}\right) \ket{\sigma_1,\,\dots,\,\sigma_N}\,,
    \label{eq: MPS scar rep}
\end{equation}
where the matrices $A^{\sigma}$ are defined as 
\begin{equation}
    A^\sigma = \begin{pmatrix}
        \cosh{\frac{\beta}{2}}& (-1)^\sigma\sqrt{\frac{\sinh(\beta)}{2}}\\
        (-1)^\sigma\sqrt{\frac{\sinh(\beta)}{2}}&\sinh{\frac{\beta}{2}}
    \end{pmatrix}\,.
    \label{eq: mps matrices}
\end{equation}
 Furthermore, the state (equation (9) in the main text)
 can also be represented as
\begin{equation}
    \ket{\bar{\mc S}(\beta)} = \frac{1}{\mathcal{Z}}\sum_{\left\{s_i=\pm1\right\}}\exp\left(\beta\sum_i s_is_{i+1} \right)\ket{\left\{s_i\right\}}\,,
\end{equation}
where the normalization factor $\mathcal{Z}$ has the interpretation of a classical partition function with classical spin variable $s_i$ at each site. In this particular case, it is simply the one for the classical Ising model, i.e., $\mathcal{Z} = \sum_{\left\{s\right\}}e^{\beta\sum_i s_is_{i+1}}$. States allowing this kind of description are known in the literature as \textit{square-root states}, since they can be written compactly as $\psi(\beta) = \sqrt{e^{-\beta h}/\mathcal{Z}}$ \cite{Swingle2016}.

The physics described by this state is simple. The average of $Z$ operators always vanishes $\left<Z_i\right>=0$, but it can develop long-range correlations as the parameter $\beta$ gets larger. 
With the matrix product (MPS) representation, we can easily compute its correlations, for which we obtain for $\mc C_{zz}(\beta)$
\begin{align}
    \mc C_{zz}(\beta) &= \frac{\bra{\bar{\mc S}(\beta)}Z_{i}Z_{i+r}\ket{\bar{\mc S}(\beta)}}{\braket{\bar{\mc S}(\beta)|\bar{\mc S}(\beta)}}\,,\nonumber\\
    &=\frac{\tanh{(\beta)}^{N-r}+\tanh{(\beta)}^r}{1+\tanh{(\beta)}^N}\,,
\end{align}
and $\mc C_{xx}(\beta)$
\begin{align}
  \mc C_{xx}(\beta) &= \frac{\bra{\bar{\mc S}(\beta)}X_{i}X_{i+r}\ket{\bar{\mc S}(\beta)}}{\braket{\bar{\mc S}(\beta)|\bar{\mc S}(\beta)}}\,,\nonumber\\
&= \frac{\cosh{(\beta)}^{N-4}}{\cosh{(\beta)}^N+\sinh{(\beta)}^N}\,,
\end{align}
which clearly does not depend on the distance $r$ between the spins, and therefore, can be factorized into single spin correlations. 
\subsection{Degenerate QMBS subspace \label{sec: ferro scars}} 
Similarly, the other Hamiltonian we are interested into examining is given by
\begin{equation}
    H_\text{D}(\beta) = \sum_i \Bar{\alpha}_i \left(e^{-\beta\left(X_i+X_{i+1}\right)}-Z_iZ_{i+1}\right)\,,
    \label{eq: ferro hamiltonian}
\end{equation}
where $\alpha_i=|\alpha|+(-1)^i$, similarly as in the nondegenerate case; this Hamiltonian is in the SMF-decomposable class and is also an example of a non-integrable Hamiltonian, as indicated by its level spacing statistics Fig.(\ref{fig:entanglement-ferro-scar+lvlspacingferro})

It is not difficult to check that the pair of states
\begin{equation}
    \ket{\mc S_a(\beta)} = e^{\frac{\beta}{2}\sum_i X_i}\otimes_i \ket{a}_i\,, ~~~a=\left\{0,1\right\}\,,
    \label{eq: Supp ferro scars}
\end{equation}
are annihilated by the Hamiltonian \eqref{eq: ferro hamiltonian} and represent a degenerate scar subspace at zero energy, as it can be seen from the entanglement entropy plot in Fig.(\ref{fig:entanglement-ferro-scar+lvlspacingferro}). 

These scar states $\ket{\mc S_a}$ have exactly zero entanglement entropy, a fact that can be seen by noticing that the states \ref{eq: Supp ferro scars} are simple product states of the form
\begin{equation}
    \ket{\mc S_a(\beta)} = \otimes_i\ket{\psi_i}\,, ~\ket{\psi}_i = \cosh{\frac{\beta}{2}}\ket{a}+\sinh{\frac{\beta}{2}}\ket{a+1}\,,
\end{equation}
where the sum $a+1$ is defined $\mod2$. It is quite simple to check that these states are orthogonal, since 
\begin{equation}
    \braket{\mc S_1(\beta)|\mc S_2(\beta)} = (\tanh{\beta})^N \underset{\text{Large}\,N}{\rightarrow} 0\,,
\end{equation}
and their average values $\braket{Z}$ and $\braket{X}$ are simply
\begin{equation}
    \braket{Z} = (-1)^a\frac{1}{\cosh{\beta_1}}\,, ~~~\braket{X} = \tanh{\beta}\,.
\end{equation}
Furthermore, notice that for $\beta=0$, we recover the standard results for the typical ferromagnetic states. 
\section{Matrix Product State representation of scars states \label{app: C}}
For completeness, we give a brief description of results needed to obtain the MPO representation of the exponential operators appearing on the states $\ket{\mc S_a}$ and $\ket{\bar{S}}$. which allows us to obtain their representations as matrix product states. Consider exponential $\exp(\frac{\beta}{2} Z\otimes Z)$, which can be expressed as
\begin{align}
    e^{\frac{\beta}{2} \,Z\otimes Z} = \cosh\left(\frac{\beta}{2}\right) \mathbf{1}\otimes\mathbf{1}+Z\otimes Z \sinh\left(\frac{\beta}{2}\right)\,,
\end{align}
notice that this exponential can be written in a convenient form as
\begin{widetext}
    \begin{equation}
e^{\frac{\beta}{2} \,Z\otimes Z} = \begin{pmatrix}
    \sqrt{\cosh\left(\frac{\beta}{2}\right)}& 0
\end{pmatrix}\begin{pmatrix}
    \sqrt{\cosh\left(\frac{\beta}{2}\right)}\\
    0
\end{pmatrix}\mathbf{1}\otimes \mathbf{1}+\begin{pmatrix}
    0&\sqrt{\sinh\left(\frac{\beta}{2}\right)}
\end{pmatrix}\begin{pmatrix}
    0\\\sqrt{\sinh\left(\frac{\beta}{2}\right)}
\end{pmatrix}Z\otimes Z\,,
    \end{equation}
\end{widetext}
if one names $B_0 = \begin{pmatrix}
    \sqrt{\cosh\left(\frac{\beta}{2}\right)} & 0
\end{pmatrix}^T$ and $B_1=\begin{pmatrix}
    0 & \sqrt{\sinh\left(\frac{\beta}{2}\right)}
\end{pmatrix}^T$, we can then write the operator in a very convenient form as
\begin{equation}
    e^{\beta Z\otimes Z} = \sum_{i,\,j}(B^T_i B_j)\mc Z^i\otimes \mc Z^j\,,
    \label{eq: MPO rep exponential ZZ}
\end{equation}
where we have defined $\mc Z^0 = \mathbf{1}$ and $\mc Z^1 = Z$. Notice that the product $B^T_iB_j$ is a $2\times 2$ matrix. 
 
With the identity \eqref{eq: MPO rep exponential ZZ} at hand, it is straightforward to obtain the tensor network representation of the state (9) in the main text. First, notice that the product of these exponential terms can be written as 
\begin{align}
    \prod_i &e^{\frac{\beta}{2} Z_iZ_{i+1}} =\nonumber \\ \sum_{\left\{i_k,\,j_k\right\}}& \text{Tr}\left[\left(B_{j_1}B_{i_1}^T\right)\dots \left(B_{j_N}B_{i_N}^T\right)\right]\mc Z_1^{i_1+j_1}\otimes\dots Z_N^{i_N+j_N}\,,\nonumber
\end{align}
where the sum $i_k+j_k$ is understood as a binary sum. We can obtain a simplified form by introducing a change of variables $i_k+j_k=q_k$, this change allow us o write the operator in a more convenient form
\begin{equation}
    \prod_i e^{\frac{\beta}{2} Z_iZ_{i+1}} = \sum_{\left\{q_k\right\}} \text{Tr}\left[C^{q_1}\dots C^{q_N}\right]\mc Z_1^{q_1}\otimes \mc Z_N^{q_N}\,,
    \label{App: MPO rep prod ZZ}
\end{equation}
with the matrices 
\begin{align}
    C^0 &= \sum_{i} B_{i\oplus0}B_{i}^T = \begin{pmatrix}
        \cosh\left(\frac{\beta}{2}\right)&0\\
        0&\sinh\left(\frac{\beta}{2}\right)\end{pmatrix}\,,\\
        C^1&=\sum_{i} B_{i\oplus1}B_{i}^T = \begin{pmatrix}
            0& \sqrt{\frac{\sinh(\beta)}{2}}\\
            \sqrt{\frac{\sinh(\beta)}{2}}&0
        \end{pmatrix}.
\end{align}

The tensor network representation of the MPS state \eqref{eq: MPS scar rep} can be obtained by applying the MPO \eqref{App: MPO rep prod ZZ} on top of the anchor product state $\otimes_i \ket{+}_i$ once we re-write it in terms of the eigenstates of the Pauli $Z$ operator. After this procedure, we recover the representation \eqref{eq: MPS scar rep} with the $A$-matrices given by a combination of the $C$-matrices above
\begin{equation}
    A^\sigma = C^0+(-1)^\sigma C^1\,, ~~\sigma=\pm 1\,.
\end{equation}
Representing the operator $e^{\beta X_i /2}$ is trivial, since it is local. The same holds for its product along the chain. 

\section{Entanglement Profile and fidelity loss \label{app: Fidelity}}

To compute the fidelity numerically at finite system sizes, we examine the behavior of the bipartite entanglement entropy. In the limits $\lambda \to 0$ and $\lambda \to \infty$, the model hosts exact scar states at zero energy. Motivated by this, we search for states with anomalously low entanglement entropy within a finite energy window of excited states. For clarity, we display this procedure explicitly in Fig.\ref{fig:entanglement-fidelity}. As the perturbation $\lambda$ increases, the entanglement of the scar states grows until these states become indistinguishable from the thermal bulk of the entanglement profile. The data shown in Fig.\ref{fig:entanglement-fidelity} complements the fidelity-loss results discussed in Sec.\ref{sec: fidelity loss}. We can observe that as we increase the coupling both states remain at relatively small entropy in comparison to the rest of the states at nearby energies, then the antisymmetric scar (in the figure labeled as scar 2) thermalizes first, going up in entropy. Meanwhile, the symmetric scar (labeled as scar 1) survives a bit longer, but eventually hybridizes as the perturbation gets larger.For $\lambda>0$, a two-fold degenerate subspace (highlighted by the blue circles) persists due to the preserved $\mathbb{Z}_2$ symmetry. However, the two states in this subspace exhibit different entanglement entropies. Notably, at $\lambda = 0.20$, one scar state remains identifiable, whereas for $\lambda \gtrsim 0.22$ no low-entanglement states are found. This transition aligns with the sharp drop in fidelity shown in Fig.\ref{Fig: Fid_loss_theory}.

\end{document}